\documentclass[referee]{raa}
\usepackage{natbib}
\usepackage{amssymb,amsmath}
\bibpunct{(}{)}{;}{a}{}{,}
\usepackage{graphicx}
\usepackage{dcolumn}
\usepackage{bm}
\usepackage[pagebackref=true]{hyperref}
\usepackage{longtable}

\begin{document}

\titlerunning{Test of conformal gravity}
\title{Test of conformal gravity as an alternative to dark matter from the observations of elliptical galaxies}

 \volnopage{ {\bf 20XX} Vol.\ {\bf X} No. {\bf XX}, 000--000}
   \setcounter{page}{1}

\author{Li-Xue Yue \inst{1,2} and Da-Ming Chen \inst{1,2}}

 \institute{National Astronomical Observatories, Chinese Academy of Sciences, Beijing 100101, China; {\it lxyue@nao.cas.cn}, {\it cdm@nao.cas.cn}\\
\and
School of Astronomy and Space Science, University of Chinese Academy of Sciences, Beijing
100049, China \\
\vs\no
   {\small Received 20xx month day; accepted 20xx month day}}

\abstract
{As an alternative gravitational theory to General Relativity (GR), the Conformal Gravity (CG) has recently been successfully verified by observations of Type Ia supernovae (SN Ia) and the rotation curves of spiral galaxies. The observations of galaxies only pertain to the non-relativistic form of gravity. In this context, within the framework of the Newtonian theory of gravity (the non-relativistic form of GR), dark matter is postulated to account for the observations. On the other hand, the non-relativistic form of CG predicts an additional potential: besides the Newtonian potential, there is a so-called linear potential term, characterized by the parameter $\gamma^*$, as an alternative to dark matter in Newtonian gravity. To test CG in its non-relativistic form, much work has been done by fitting the predictions to the observations of circular velocity (rotation curves) for spiral galaxies. In this paper, we test CG with the observations from elliptical galaxies. Instead of the circular velocities for spiral galaxies, we use the velocity dispersion for elliptical galaxies. By replacing the Newtonian potential with that predicted by non-relativistic form of CG in Hamiltonian, we directly extend the Jeans equation derived in Newtonian theory to that for CG. By comparing the results derived from the ellipticals with that from spirals, we find that the extra potential predicted by CG is not sufficient to account for the observations of ellipticals. Furthermore, we discover a strong correlation between $\gamma^*$ and the stellar mass $M^*$ in dwarf spheroidal galaxies. This finding implies that the variation in $\gamma^*$ violates a fundamental prediction of Conformal Gravity (CG), which posits that $\gamma^*$ should be a universal constant.}
\maketitle

\section{Introduction}
Einstein's General Relativity (GR) has been verified very successfully on the scale of the solar system, where the vacuum solutions of Einstein's equation, known as Schwarzschild metric, are applied. On larger scales, in particular when it comes to the studies of galaxies and cosmology, dark matter (DM) and dark energy (DE) are assumed to account for observations. Since both DM and DE lack direct theoretical supports and observational evidence, many efforts are devoted to the modified gravity alternative to GR and its non-relativistic form, Newtonian gravity. For~instance, one can enhance the standard Lagrangian in general relativity by incorporating higher-order curvature corrections~\citep{lovelock1971einstein,Lovelock1972,PhysRevLett.55.2656,2005GReGr..37.1869K,Oikonomou_2021,brassel2022charged}, or~formulate non-linear Lagrangians~\citep{buchdahl1970non,goswami2014collapsing}. Other relevant examples include modified Newtonian dynamics (MOND)~\citep{1983ApJ...270..365M,famaey:hal-02927744} and its relativistic version~\citep{2004PhRvD..70h3509B}, conformal gravity ~\citep{Mannheim_1997,MANNHEIM2006340}, as well as the quantum effects on cosmic scales as an alternative to dark matter and dark energy~\citep{Chen_2022,universe10080333}. Clearly, any modifications or extensions to GR should be verified by observations, in particular by the observations from the solar system. However, in the solar system, the effects of any modifications or extensions to GR should be negligible since on this system scale GR turns out to be exact when predicting observations. On galactic scales, the non-relativistic theory of gravity suffices. For Newtonian theory, DM is introduced to produce extra gravitational potential so that when combined with the potential created by the luminous matter, the total gravitational potential can account for the observations of galaxies. On the other hand, in any modified theory of gravity, it is required that, besides the usual Newtonian potential, the luminous matter must produce extra gravitational potential to replace the potential produced by DM in Newtonian theory. 

In recent years, the Conformal Gravity (CG) has attracted much interest in testing it as an alternative to DM and DE with astronomical observations (for a review see~\cite{MANNHEIM2006340}). As a relativistic theory alternative to GR, CG can solve the lang-standing cosmological constant problem encountered in standard $\Lambda$CDM cosmological model\citep{1992ApJ...391..429M,Mannheim2000FoPh,Mannheim_2001}, and the CG cosmology has been tested with SN Ia data~\citep{MANNHEIM2006340,YANG201343} . In its non-relativistic limit, luminous matter generates additional gravitational potential beyond the conventional Newtonian potential~\citep{1989ApJ...342..635M}. This could potentially resolve the missing mass problem observed in galaxies and galaxy clusters without the need for DM. To assess Conformal Gravity (CG) in its non-relativistic form, a significant amount of research has been conducted. This involved fitting the theoretical predictions to the observed circular velocities (rotation curves) of spiral galaxies~\citep{mannheim_fitting_2012,Mannheim_2013,2015JPhCS.615a2002O}.

In this paper, we take a different approach. We test CG using the observations from elliptical galaxies. Instead of relying on the circular velocities characteristic of spiral galaxies, we utilize the velocity dispersion of elliptical galaxies. Specifically, in the Hamiltonian, we substitute the Newtonian potential with the one predicted by the non-relativistic form of CG. By doing so, we directly extend the Jeans equation, which was originally derived within the framework of Newtonian theory, to the context of CG.

The remainder of this paper is structured as follows. In Section~\ref{sec:review CG}, we review the fundamentals of Conformal Gravity (CG) and present the necessary formulas. In Section~\ref{sec:test of CG}, we first give a brief introduction to the test of CG using spiral galaxies. Subsequently, we elaborate in detail on the procedures we adopted when applying CG to elliptical galaxies. The conclusions and discussions are presented in Section~\ref{sec:conclusions}.

\section{conformal gravity}\label{sec:review CG}
 In comparison to General Relativity (GR), Conformal Gravity (CG) is formulated by maintaining the metric as the gravitational field. However, it endows gravity with an additional symmetry, namely the conformal symmetry, which extends beyond the ordinary coordinate invariance. By imposing the principle of local conformal invariance as the requisite principle to restrict the choice of action for the gravitational field in curved spacetime, one requires the uniquely selected fourth-order gravitational action~\citep{1989ApJ...342..635M}
\begin{equation}\label{eq:Weyl action}
\begin{aligned}
    I_W =& -{\alpha}_g \int d^4x \sqrt{-g} C_{\lambda\mu\nu\kappa} C^{\lambda\mu\nu\kappa} \\
        =& -\alpha_g\int d^4x\sqrt{-g}\left[R_{\lambda\mu\nu\kappa}R^{\lambda\mu\nu\kappa}-2R_{\mu\nu}R^{\mu\nu}+(1/3){(R^\alpha_{\phantom{\alpha}\alpha})}^2\right] \\
        =& -2{\alpha}_g \int d^4x \sqrt{-g}\left[R_{\mu\nu}R^{\mu\nu}-(1/3){(R^\alpha_{\phantom{\alpha}\alpha})}^2\right] 
\end{aligned}
\end{equation}
 to remain invariant under any local metric transformation $g_{\mu \nu}(x) \to e^{2\alpha(x)} g_{\mu \nu}(x)$ (called conformal transformation), and thus an action satisfying conformal symmetry. In Equation (\ref{eq:Weyl action}), $\alpha_g$ is a dimensionless coupling constant, and $C^{\lambda\mu\nu\kappa}$ is the conformal Weyl tensor defined by~\citep{MANNHEIM2006340}
 \begin{equation}\label{eq:Weyl tensor}
    C_{\lambda\mu\nu\kappa} = R_{\lambda\mu\nu\kappa} - \frac{1}{2} \left(g_{\lambda\nu} R_{\mu\kappa} - g_{\lambda\kappa} R_{\mu\nu} - g_{\mu\nu} R_{\lambda\kappa} +
                              g_{\mu\kappa} R_{\lambda\nu}\right) + \frac{1}{6} R^{\alpha}_{\phantom{\alpha}\alpha} \left(g_{\lambda\nu} g_{\mu\kappa} - g_{\lambda\kappa} g_{\mu\nu}\right),
 \end{equation}
i.e., a tensor constructed by a particular combination of the Riemann and Ricci tensors and the Ricci scalar. The particular property of Weyl tensor is that it has the kinematic relation $g_{\mu\kappa}C^{\lambda_{\mu\nu\kappa}}=0$. In other words, Weyl tensor is traceless. 

Conformal gravity requires the energy-momentum tensor $T_{\mu\nu}$ to be traceless, i.e., $T^{\mu}_{ \mu}=0$. On the other hand, elementary particle masses are not kinematic, but rather that they are acquired dynamically by spontaneous breakdown. Hence, consider a massless, spin-$\frac{1}{2}$ matter field fermion $\psi(x)$ which is to get its mass through a massless, real spin-0 Higgs scalar boson field $S(x)$. The required matter field action $I_M$ can be defined by ~\citep{mannheim_fitting_2012}
\begin{equation}
  I_M = - \int d^4x \sqrt{-g} \left[ \frac{1}{2} S^{;\mu} S_{;\mu} - \frac{1}{12} S^2 R^\mu_{\phantom{\mu} \mu} + \lambda S^4 + i \bar{\psi} \gamma^\mu(x) \left( \partial_\mu + \Gamma_{\mu}(x) \right) \psi - h S \bar{\psi} \psi \right],
\end{equation}
where $h$ and $\lambda$ are dimensionless coupling constants, $\gamma^{\mu}(x)$ are the Dirac matrices and $\Gamma_{\mu}(x)$ are the fermion spin connection. 
Variation of $I_M$ with respect to the metric yields the energy-momentum tensor
\begin{equation}\label{eq:general energy momentum tensor}
  \begin{aligned}
      T^{\mu\nu} =\ & i\bar{\psi}\gamma^{\mu}(x)[\partial^\nu + \Gamma^{\nu}(x)]\psi + \frac{2}{3}S^{;\mu}S^{;\nu} - \frac{1}{6}g^{\mu\nu}S^{;\alpha}S_{;\alpha} - \frac{1}{3}S S^{;\mu;\nu} + \frac{1}{3}g^{\mu\nu}S S^{;\alpha}_{\phantom{;\alpha};\alpha} \\
      & - \frac{1}{6}S^2\left(R^{\mu\nu} - \frac{1}{2}g^{\mu\nu}R^{\alpha}_{\phantom{\alpha}\alpha}\right) - g^{\mu\nu}\left[\lambda S^4 + i\bar{\psi}\gamma^{\alpha}(x)[\partial_{\alpha}+\Gamma_{\alpha}(x)]\psi - hS\bar{\psi}\psi\right].    
  \end{aligned}
  \end{equation}
 
The total action is $I=I_W+I_M$,  variation of the total action  with respect to the metric then yields~\citep{MANNHEIM2006340}
\begin{equation} \label{eq:eom}
    \frac{1}{{(-g)}^{1/2}} \frac{\delta I}{\delta g_{\mu \nu}} = -2{\alpha}_g W^{\mu \nu} +\frac{1}{2} T^{\mu \nu} = 0, 
\end{equation}
where $W^{\mu \nu} = \left[ W^{\mu \nu}_{(2)} - \frac{1}{3} W^{\mu \nu}_{(1)} \right]$, and
\begin{equation}\label{eq:sijiezhangliangfenshi}
    \begin{aligned}
        W^{\mu\nu}_{(1)} &= 2 g^{\mu\nu} (R^{\alpha}_{\phantom{\alpha}\alpha})^{;\beta}_{\phantom{;\beta};\beta} - 2 (R^{\alpha}_{\phantom{\alpha}\alpha})^{;\mu;\nu} - 2 R^{\alpha}_{\phantom{\alpha}\alpha} R^{\mu\nu} + \frac{1}{2} g^{\mu\nu} (R^{\alpha}_{\phantom{\alpha}\alpha})^2, \\
        W^{\mu\nu}_{(2)} &= \frac{1}{2} g^{\mu\nu} (R^{\alpha}_{\phantom{\alpha} \alpha})^{;\beta}_{\phantom{;\beta};\beta} + (R^{\mu\nu})^{;\beta}_{\phantom{;\beta};\beta} - (R^{\mu\beta})^{;\nu}_{\phantom{;\beta};\beta} - (R^{\nu\beta})^{;\mu}_{\phantom{;\beta} ;\beta} - 2 R^{\mu\beta} R^{\nu}_{\phantom{\beta}\beta} + \frac{1}{2} g^{\mu\nu} R^{\alpha\beta} R_{\alpha\beta}.
    \end{aligned} 
\end{equation}

\subsection{Applying to cosmology}
In applying conformal gravity to cosmology, Weyl tensor vanishes in a Robertson-Walker metric~\citep{1992ApJ...391..429M}
\begin{equation}
    ds^2 = c^2 dt^2 - R^2(t) \left[ \frac{dr^2}{1 - Kr^2} + r^2 d\theta^2 + r^2 \sin^2 \theta d\phi^2 \right].
\end{equation}
Thus $W^{\mu\nu}=0$, and we see from Equation (\ref{eq:eom}) that $T^{\mu\nu}=0$. It turns out that conformal symmetry forbids the presence of any fundamental cosmological term, and is thus a symmetry which is able to control the cosmological constant. Even after the conformal symmetry is spontaneously broken (as is needed to generate particle mass), the contribution of an induced cosmological constant to cosmology will still be under control~\citep{MANNHEIM2006340}. Consequently, CG is potentially capable of solving the cosmological constant problem. The full content of the theory can be obtained by choosing a particular gauge in which the scalar field takes the constant value $S_0$. In this case, the energy-momentum tensor of Equation (\ref{eq:general energy momentum tensor}) becomes~\citep{MANNHEIM2006340,MANNHEIM2017125}
\begin{equation}\label{eq:cos energy momentum tensor}
  T^{\mu\nu} = i\bar{\psi}\gamma^{\mu}(x)\left[\partial^\nu + \Gamma^{\nu}(x)\right]\psi - \frac{1}{6} S_0^2\left(R^{\mu\nu} - \frac{1}{2}g^{\mu\nu}R^{\alpha}_{\phantom{\alpha}\alpha}\right) - g^{\mu\nu}\lambda S_0^4 = 0.
\end{equation}
An averaging of $i\bar{\psi}_{\mu}(x)[\partial + \Gamma_{\nu}(x)]\psi$ over all the fermionic modes propagating in a Robertson-Walker background will bring the fermionic contribution to $T^{\mu\nu}$ to the form of a kinematic perfect fluid
\begin{equation}\label{eq:perfect fluid}
T^{\mu\nu}_{\text{kin}} = \frac{1}{c} \left[(\rho_m + p_m) U^\mu U^\nu + p_m g^{\mu\nu}\right],
\end{equation}
thus the conformal cosmology equation of motion can be written as~\citep{MANNHEIM2006340}
\begin{equation}\label{eq:equation of motion 1}
    \frac{1}{6} S_0^2 \left(R^{\mu\nu} -  \frac{1}{2} g^{\mu\nu} R^\alpha_{\phantom{\alpha} \alpha} \right) 
        = \frac{1}{c} \left[(\rho_m + p_m) U^\mu U^\nu + p_m g^{\mu\nu}\right] - g^{\mu\nu} \lambda S_0^4.
\end{equation}
Comparing with the standard Einstein equation, we only need to replace the gravitational constant $G$ by an effective, dynamically induced one $G_{\text{eff}} = - 3c^3/(4\pi S_0^2)$~\citep{1992ApJ...391..429M}. We define conformal analogs of the standard $\Omega_M(t)$,  $\Omega_{\Lambda}(t)$ and $\Omega_K(t)$ via
\begin{equation}
    \bar{\Omega}_M(t) = \frac{8\pi G_{\text{eff}} \rho_m(t)}{3c^2 H^2(t)}, \quad \bar{\Omega}_\Lambda(t) = \frac{8\pi G_{\text{eff}} \Lambda}{3c H^2(t)}, 
                        \quad \bar{\Omega}_K(t) = -\frac{K c^2}{R^2(t) H^2(t)},
\end{equation}
where $H(t)=\dot{R(t)}/R(t)$ is the Hubble parameter and $\Lambda = \lambda S_0^4$. As usual, in a Robertson-Walker geometry Equation (\ref{eq:equation of motion 1}) yields, at redshift $z$, the expression of the Hubble parameter 
\begin{equation}
    H(z) = H_0 \sqrt{\bar{\Omega}_{M} {(1+z)}^3 + \bar{\Omega}_{K} {(1+z)}^2 + \bar{\Omega}_{\Lambda}}
\end{equation}
where $\bar{\Omega}_{M}=\bar{\Omega}_{M}(t=0)$, and so on. In subsequent calculations, we adopt the values  $\bar{\Omega}_{K}=0.67$, $\bar{\Omega}_{\Lambda }=0.33$, and H$_0$ = 69.3 \,km s$^{-1}$  Mpc$^{-1}$, as per reference ~\citep{YANG201343}. 

For future reference, we define the angular diameter distance as
\begin{equation}\label{eq:angular diameter distance}
    D_\text{A}(z_1,z_2)=\frac{1}{1+z_2}f_K\left[\chi(z_1,z_2)  \right],\quad f_K(\chi) = (-K)^{-1/2}\sinh\left[(-K)^{1/2}\chi\right],
\end{equation}
where
\begin{equation}
  \chi(z_1,z_2) = \int_{z_1}^{z_2} \frac{cdz'}{H(z')}.
\end{equation}

\subsection{Non-relativistic limit}
To conduct a test of Conformal Gravity (CG) using galaxy observations, it is necessary to derive the non-relativistic limit of CG. Mannheim and Kazanas ~\citep{1989ApJ...342..635M, mannheim_newtonian_1994} found an exact CG analog of the Schwarzschild exterior and interior solutions to standard gravity by solving the equation $4\alpha_gW^{\mu\nu}=T^{\mu\nu}$ for a static, spherically symmetric source. It turns out that the full kinematic content of CG is contained in the line element~\citep{MANNHEIM2006340}
\begin{equation}\label{eq:pushiCGdugui}
    ds^2 = -B(r) dt^2 + \frac{dr^2}{B(r)} + r^2 (d\theta^2 + \sin^2 \theta d\phi^2).   
\end{equation}
Evaluating the form that $W^{\mu\nu}$ takes in this line element leads to
\begin{equation}\label{eq:Wrr}
\begin{aligned}
    \frac{W^{rr}}{B(r)} = &\frac{B' B'''}{6} - \frac{(B'')^2}{12} - \frac{1}{3r}(B B''' - B' B'')  \\ 
                          & - \frac{1}{3r^2}(B B'' + B'^2) + \frac{2 B B'}{3r^3} - \frac{B^2}{3r^4} + \frac{1}{3r^4}
\end{aligned}
\end{equation}
and
\begin{equation}\label{eq:W00}
  \begin{aligned}
    W^{00} = &-\frac{B''''}{3} + \frac{(B'')^2}{12B} - \frac{B''' B'}{6B} - \frac{B'''}{r} - \frac{B'' B'}{3rB}  \\ 
             &+ \frac{B''}{3r^2} + \frac{(B')^2}{3r^2 B} - \frac{2 B'}{3r^3} - \frac{1}{3r^4 B} + \frac{B}{3r^4}
\end{aligned}
\end{equation}
for its components of interest. Combining Equations (\ref{eq:Wrr}) and (\ref{eq:W00}) then yields
\begin{equation}\label{eq:combined solution}
    \frac{3}{B}(W^0_{\phantom{0}0} - W^r_{\phantom{r}r}) = B'''' + \frac{4 B'''}{r} = \frac{1}{r}(rB)'''' = \nabla^4 B.
\end{equation}
It is convenient to define a source function $f(r)$ via
\begin{equation}\label{eq:f(r)}
    f(r) = \frac{3}{4\alpha_g B(r)} \left(T^0_{\phantom{0}0} - T^r_{\phantom{r}r} \right)
\end{equation}
so that the equations of motion of Equation (\ref{eq:eom}) can be written 
\begin{equation}\label{eq:jianhuayundongfangcheng}
    \nabla^4 B(r) = f(r).
\end{equation}
We are interested in the exterior solution to Equation (\ref{eq:jianhuayundongfangcheng}) for a static, spherically source of radius $r_0$, which is readily given by
\begin{equation}\label{eq:exterior solution}
    B(r > r_0) = -\frac{r}{2} \int_0^{r_0} dr' \, r'^2 f(r') - \frac{1}{6r} \int_0^{r_0} dr' \, r'^4 f(r') + w - \kappa r^2, 
\end{equation}
where $w - \kappa r^2$ term is the general solution to the homogeneous equation $\nabla^4 B(r) =0$. On defining 
\begin{equation}\label{eq:mkcanshu}
    \gamma = -\frac{1}{2} \int_0^{r_0} dr' \, r'^2 f(r') , \qquad 2\beta = \frac{1}{6} \int_0^{r_0} dr' \, r'^4 f(r'),
\end{equation}
dropping $kr^2$ term and setting $w=1$, the metric of Equation (\ref{eq:exterior solution}) can be written, without any approximation, as
\begin{equation}\label{eq:final metric exterior to r0}
    B(r>r_0) = -g_{00}=\frac{1}{g_{rr}}=1 -  \frac{2\beta}{r}  + \gamma r.
\end{equation}
The Schwarzschild-like vacuum solutions of any modified theory of gravity offer us an opportunity to verify the theory in its non-relativistic form. Specifically, this verification can be carried out on the scales of solar systems, galaxies, and galaxy clusters. In such scenarios, the metric $g_{\mu\nu}$ is reduced to gravitational potential $V$. In terms of gravitational potential $V(r)$, we can rewrite the metric of Equation (\ref{eq:final metric exterior to r0}) as
\begin{equation}\label{eq:metric in terms of V}
 B(r>r_0)=1+2V(r)/c^2, \ \ \text{with} \ \ V(r)=V_{\beta}+V_{\gamma} \ \ \text{and} \ \ V_{\beta}=-\frac{\beta c^2}{r}, \ \  V_{\gamma}=\frac{1}{2}\gamma c^2 r.
\end{equation}
In the region where $2\beta/r\gg\gamma r$, when $\beta=GM/c^2$, the Schwarzschild solution $B(r>r_0)=1-\frac{2GM}{c^2r}$ can be recovered. Departures from this solution, specifically the linear potential $V_{\gamma}=\gamma c^2 r/2$, only occur at large distances. As a result, the standard solar system Schwarzschild phenomenology is preserved. 

\section{Test of conformal gravity with observations of galaxies}\label{sec:test of CG}
As previously shown, when verifying a new relativistic theory of gravity through galaxy observations, one must transition from the geometric perspective (utilizing the metric $g_{\mu\nu}$) to that of Newtonian dynamics (employing the gravitational potential $V$). Consequently, in the realm of galactic dynamics, the kinematic aspects are determined by the gravitational potential. This holds true regardless of the form the potential assumes and its origin. 
The potential shown in Equation (\ref{eq:metric in terms of V}) represents the potential generated by a point mass $M$ of the luminous matter in CG. Besides the conventional Newtonian potential $V_{\beta}=-\frac{\beta c^2}{r}=-GM/r$, there is also a linear potential, $V_{\gamma}=\frac{1}{2}\gamma c^2 r$. This linear potential is proposed as an alternative to the potential generated by dark matter in Newtonian theory and thus requires verification through observations of galaxies. For a typical star of the solar mass $M_{\sun}$, we write its potential as
\begin{equation}\label{eq:potential of the sun}
V^*(r)=-\frac{\beta^* c^2}{r} + \frac{\gamma^* c^2 r}{2},
\end{equation}
where $\beta^* = GM_\odot/c^2 = 1.48 \times 10^3 m$ and $\gamma^*$ can be determined by observations. If we denote $N^*=\frac{M}{M_{\sun}}$, $\beta=N^*\beta^*$ and $\gamma=N^*\gamma^*$, then for any point mass $M$, the expression for its potential shown in Equation (\ref{eq:metric in terms of V}) can be rewritten as
\begin{equation}\label{eq:potential of any mass M}
V(r)=V_{\beta}+V_{\gamma}=-\frac{N^*\beta^* c^2}{r} + \frac{N^*\gamma^* c^2 r}{2}.
\end{equation}
\subsection{Test with spiral galaxies}
Up to now, the value of $\gamma^*$ in Equation (\ref{eq:potential of any mass M}) has been uniquely determined  by the rotation curve observations of spiral galaxies. For example, the circular velocity $v_c(R)$ contributed by luminous matter in the equatorial plane $z=0$ of a axisymmetric disk galaxy of surface mass density $\Sigma(R)$ is 
\begin{equation}\label{eq:circular velocity for luminous matter}
v_c^2(R)=R\frac{d V_{\text{LOC}}(R)}{d R}=RV'_{\beta}(R)+RV'_{\gamma},
\end{equation}
where $V_{\text{LOC}}(R)=V_{\beta}(R)+V_{\gamma}(R)$, with~\citep{MANNHEIM2006340}
\begin{equation}\label{eq:Vbeta(R,z)}
\begin{aligned}
V_{\beta}(R,z)=& -N^*\beta^*c^2\int\frac{d M}{|\vec{r}-\vec{r'}|} \\
              =& -N^*\beta^*c^2\int_{0}^{\infty}d R'\int_{0}^{2\pi}d\phi'\int_{-\infty}^{\infty}d z'\frac{R'\Sigma(R')\delta(z')}{(R^2+R'^2-2RR'\cos\phi'+(z-z')^2)^{1/2}} \\
              =& -2\pi N^*\beta^*c^2\int_{0}^{\infty}d k\int_{0}^{\infty}d R' R'\Sigma(R')J_0(kR)J_0(kR')e^{-k|z|}.
\end{aligned}
\end{equation}
and
\begin{equation}\label{eq:Vgamma(R,z)}
\begin{aligned}
V_{\gamma}(R,z)=& \frac{N^*\gamma^*c^2}{2}\int d M |\vec{r}-\vec{r'}| \\
               =& \frac{N^*\gamma^*c^2}{2}\int_{0}^{\infty}d R'\int_{0}^{2\pi}d\phi'\int_{-\infty}^{\infty}d z' \\
                &\Sigma(R')\delta(z') (R^2+R'^2-2RR'\cos\phi'+(z-z')^2)^{1/2} \\
              =& \pi N^*\gamma^*c^2\int_{0}^{\infty}d k\int_{0}^{\infty}d R' \\
               & R'\Sigma(R')[(R^2+R'^2)J_0(kR)J_0(kR')-2RR'J_1(kR)J_1(kR')]e^{-k|z|}.
\end{aligned}
\end{equation}
 
However, when fitting to the rotation curves of spiral galaxies, it has been found that there exists a universal, galaxy-independent linear potential, $V_{\gamma_0}=\frac{1}{2}\gamma_0 c^2r$. This potential can be ascribed to the effect of the potentials due to the rest of matter in the universe on any local galaxies~\citep{Mannheim_1997}. Consequently, around a point mass $M$, the total potential on a test particle is
\begin{equation}\label{eq:total potential}
    V(r) = V_\text{LOC}(r)+\frac{\gamma_0c^2r}{2},  \text{with} \ \  V_\text{LOC}(r)=-\frac{N^*\beta^* c^2}{r} + \frac{N^*\gamma^* c^2 r}{2},    
\end{equation}
By fitting rotation curves of spiral galaxies~\citep{MANNHEIM2006340}, it is found that 

\begin{equation}\label{eq:value of gammas}
    \gamma^\ast=5.42 \times 10^{-39} m^{-1}, \, \gamma_0=3.06 \times 10^{-28} m^{-1}.
\end{equation}
In what follows, we will determine $\gamma^*$ and $\gamma_0$ via a different approach, namely by using the observations of elliptical galaxies.

\subsection{Test with elliptical galaxies:theory}
The observable quantities of elliptical galaxies that we can utilize are the surface brightness and velocity dispersion. To validate Conformal Gravity (CG) using these observations, we begin with the Jeans equation for static gravitational systems. Generally speaking, for static systems, the modified Hamiltonian of any new gravitational theory can be straightforwardly constructed by the replacement of the Newtonian potential $V_N(\vec{x})$ in $H=\frac{1}{2}v^2+V_N(\vec{x})$ with the modified potential $V(\vec{x})$. In this paper, we make use of the potential presented in Equation (\ref{eq:total potential}). The collisionless Boltzman equation (CBE) is ~\citep{binney2011galactic} 
\begin{equation}\label{eq:CBE}
\frac{\partial f}{\partial t}+[f,H]=0,
\end{equation}
where $f$ is the distribution function (DF) in phase space $(\vec{x},\vec{v})$, and the square bracket is a Poisson bracket. In terms of inertial Cartesian coordinates, in which $H=\frac{1}{2}v^2+V(\vec{x})$, the CBE for a static system is
\begin{equation}\label{eq:CBE for static systems}
\vec{v}\cdot\nabla f-\nabla V(\vec{x})\cdot\nabla_{\vec{v}}f=0,
\end{equation}
where $\nabla_{\vec{v}}$ is the gradient in velocity $\vec{v}$ space. Jeans equation is derived from the CBE of Equation (\ref{eq:CBE for static systems}), and for static, spherical systems, it reads~\citep{binney2011galactic}
\begin{equation}\label{eq:Jeans equation}
    \frac{1}{\rho} \frac{d}{dr} (\rho \sigma_r^2) + 2 \frac{\beta(r) \sigma_r^2}{r}  = - \frac{d V}{dr},
\end{equation}
where $\rho$ is the matter density, $\sigma_r^2$ is the radial velocity dispersion, and $\beta(r)$ is the anisotropy parameter (not confused with the $\beta$ potential). Note that the gravitational potential $V$ in Equation (\ref{eq:Jeans equation}) is the one for CG, as shown in Equation (\ref{eq:total potential}).  For simplicity, we assume that the systems are  isotropic ($\beta=0$) and the velocity dispersion $\sigma_r^2=\sigma_*^2$ ($\sigma_*$ is the line of sight dispersion) is a constant for each system. Thus the Jeans equation is simplified as
\begin{equation}\label{eq:simplified Jeans equation}
    \frac{\sigma_\ast^2}{\rho(r)}  \frac{d\rho(r)}{dr} = - \frac{d V}{dr} .
\end{equation}
In the subsequent analysis, we will determine the values of $\gamma^*$ and $\gamma_0$ in the potential $V$ of Equation (\ref{eq:simplified Jeans equation}) using the data of elliptical galaxies and compare the results with those of Equation (\ref{eq:value of gammas}), which were obtained from the data of spiral galaxies. This will be achieved using the observables  $\rho(r)$ (obtained from the surface brightness $I(r)$) and $\sigma_*$ for a sample consisting of dwarf elliptical galaxies and other samples consisting  of bright elliptical galaxies.
 
 What we actually observed is the surface brightness $I(R)$, so we must extract $\rho(r)$ from it. For dwarf spheroidal galaxies, we employ the Plummer profile~\citep{Walker_2009,Moskowitz_2020}
 \begin{equation}\label{eq:Plummer surface brightness}
I(R) = L{(\pi R_e^2)}^{-1} {\left( 1 + \frac{R^2}{R_e^2} \right)}^{-2},
\end{equation}
 where $L$ is the total luminosity,  $R_e$ is the eﬀective radius, i.e., the projected radius encircling half of the total luminosity associated with $I(R)$. 
 The luminosity density $j(r)$ can be extracted from $I(R)$ via an Abel transform~\citep{binney2011galactic}
 \begin{equation}\label{eq:Abel transform}
 j(r)=-\frac{1}{\pi}\int_{r}^{\infty}dR\frac{1}{\sqrt{R^2-r^2}}\frac{dI(R)}{dR}, \ \ I(R)=2\int_{R}^{\infty}dr\frac{j(r)r}{\sqrt{r^2-R^2}}.
 \end{equation}
 By considering the mass to light ratio $\Upsilon=M/L$, we obtain the mass density for the Plummer profile
 \begin{equation}\label{eq:Plummer density}
\rho(r) = \frac{3}{4\pi} \frac{M}{R_e^3} {\left( 1 + \frac{r^2}{R_e^2} \right)}^{-5/2},
\end{equation}
 where $M$ is the total mass of the galaxy. Substituting $\rho(r)$ of Equation (\ref{eq:Plummer density}) into Equation (\ref{eq:simplified Jeans equation}) we obtain the result that can be directly used to fit the observational data for dwarf spheroidals
 \begin{equation}\label{eq:Plummer Jeans equation}
    \frac{5\sigma_\ast^2(r/R_e)^2}{r(1+r^2/R_e^2)} = \frac{d V}{dr} = V'_{\beta}+V'_{\gamma}+\frac{\gamma_0c^2}{2}.
\end{equation}
 
 For other general elliptical galaxies, we employ the Sérsic profile~\citep{1963BAAA....6...41S,1968adga.book.....S}
 \begin{equation}\label{eq:IR}
I(R) = I_0 e^{-b_n\left( R/R_e \right)^{1/n}},
\end{equation}
 where $I_0$ is the central intensity, $R_e$ is the effective radius, $n$ is the Sérsic index, and $b_n$ is the scale factor, the fitted approximate value of which is \(b_n=2n-1/3+4/405n+46/25515n^2\)~\citep{Ciotti1999AnalyticalPO}. By making use of the formula $L = 2\pi\int_0^\infty I(R') R' dR'$, one can derive the central intensity
 \begin{equation}\label{eq:central intensity}
I_0 = \frac{L b_n^{2n}}{2\pi R_e^2 n  \Gamma(2n)}.
\end{equation}
Consequently, the Sérsic density profile can be computed once more through an Abel transform of Equation (\ref{eq:Abel transform})~\citep{Prugniel1997TheFP}
\begin{equation}\label{eq:Sersic density profile}
\begin{aligned}
    \rho(r) =& \rho_0 \left( \frac{r}{R_e} \right)^{-p} \exp \left[ -b_n \left( \frac{r}{R_e} \right)^{1/n} \right],  \\
\text{with}\ \ \    \rho_0  =& \Upsilon \frac{I_0 b_n^{n (1 - p)} \Gamma(2n)}{2 R_e \Gamma(n(3 - p))},
\end{aligned}
\end{equation}
where the parameter $p$ satisfies the relationship $p = 1-1.188/2n+0.22/4n^2$. Substituting $\rho(r)$ of Equation (\ref{eq:Sersic density profile}) into Equation (\ref{eq:simplified Jeans equation}), we get the Jeans equation for the Sérsic profile as
\begin{equation}\label{eq:Jeans equation for Sersic profile}
     \left(p + \frac{b_n}{n}\left(\frac{r}{R_e}\right)^{1/n}\right) \frac{\sigma_\ast^2}{r} = \frac{d V}{dr}=V'_{\beta}+V'_{\gamma}+\frac{\gamma_0c^2}{2}.
\end{equation}

We now shift our focus to the right-hand side of Equation (\ref{eq:simplified Jeans equation}), (\ref{eq:Plummer Jeans equation}) or (\ref{eq:Jeans equation for Sersic profile}) and compute the derivatives of $V_\beta$ and $V_\gamma$. The derivative of the Newtonian potential $V_\beta$ is readily given by~\citep{MANNHEIM2006340}
\begin{equation}\label{eq:derivative of Newtonian potential}
 V'_{\beta}(r) = \frac{4\pi N^* \beta^\ast c^2}{r^2} \int_0^r dr' \, \rho(r') r'^2.
\end{equation} 
For the linear potential of the system, we have
\begin{equation}\label{eq:linear potential}
\begin{aligned}
V_\gamma(r)=&\frac{N^*\gamma^*c^2}{2}\int dM |\vec{r'}-\vec{r}| \\
           =&\pi N^*\gamma^*c^2\int_{0}^{\infty}d r'\int_{0}^{\pi}d\theta\rho(r')r'^2\sin\theta\sqrt{r^2+r'^2-2rr'\cos\theta}  \\
           =&\frac{2\pi N^*\gamma^*c^2}{3}\left[\frac{1}{r}\int_{0}^{r}d r'r'^2\rho(r')(3r^2+r'^2)+\int_{r}^{\infty}d r' r'\rho(r')(3r'^2+r^2)\right].                                                                                                     
\end{aligned}
\end{equation}
We thus obtain the derivative of the linear potential $V_\gamma$~\citep{MANNHEIM2006340}
\begin{equation}\label{eq:derivative of the linear potential}
    V'_{\gamma}(r) = \frac{2\pi N^* \gamma^\ast c^2}{3r^2} \int_0^r dr' \, \rho(r') \left( 3r^2 r'^2 - r'^4 \right) + 
        \frac{4\pi N^*\gamma^\ast c^2 r}{3} \int_r^\infty dr' \, \rho(r') r'.
\end{equation}
  
 By substituting Equations (\ref{eq:derivative of Newtonian potential}) and (\ref{eq:derivative of the linear potential}) into the right hand of the Equation (\ref{eq:Plummer Jeans equation}) for Plummer profile, we can determine $\gamma^*$ and $\gamma_0$ using the data of dwarf spheroidal galaxies. Similarly, When aiming to determine  $\gamma^*$ and $\gamma_0$ from the data of bright spheroidal galaxies, we can perform the same procedure for the Sérsic profile of the Equation (\ref{eq:Jeans equation for Sersic profile}).

On the other hand, it is intriguing to compare the results of our Conformal Gravity (CG) analysis with those predicted by the conventional dark matter model. To carry out this comparison, similar to the approach in Equation (\ref{eq:simplified Jeans equation}), we assume that the system is isotropic and the velocity dispersion remains constant. The key distinction here is that the gravitational potential $V$ follows the Newtonian form, which is generated by the combined mass of dark matter and luminous matter, denoted as dynamic mass $M_\text{dyn}$. So for Newtonian theory of gravity, Equation (\ref{eq:simplified Jeans equation}) becomes
\begin{equation}\label{eq:simplified Jeans equation for DM}
    \frac{\sigma_\ast^2}{\rho(r)}  \frac{d\rho(r)}{dr} = - \frac{d V}{dr}=\frac{GM_\text{dyn}(r)}{r^2}.
\end{equation}
Of course, this equation is valid only when we assume that mass distribution follows the light distribution. However, this assumption is generally not true because, in most cases, a significant portion of dark matter is distributed outside the region of luminous matter, forming a dark halo~\citep{Walker_2009,Moskowitz_2020}. Nevertheless, from the perspective of gravitational force, as a toy model, such a simplification can help us verify whether CG has the ability to account for the observations without invoking dark matter. Consequently, to compare the results of CG with that of Newtonian theory, we have to replace the potential in Equation (\ref{eq:Plummer Jeans equation}) and Equation (\ref{eq:Jeans equation for Sersic profile}) with Newtonian potential. Specifically, Equation (\ref{eq:Plummer Jeans equation}) is replaced by
\begin{equation}\label{eq:MJSPul}
    M_\text{dyn}(r) = \frac{5R_e\sigma_\ast^2(r/R_e)^3}{G(1+r^2/R_e^2)}
\end{equation}
 for Plummer profile assumed for dSpshs and Equation (\ref{eq:Jeans equation for Sersic profile}) is replaced by
 \begin{equation}\label{eq:MJS}
    M_\text{dyn}(r) =  \left(p + \frac{b_n}{n}\left(\frac{r}{R_e}\right)^{1/n}\right) \frac{\sigma_\ast^2 r}{G}
\end{equation}
for Sersic profile.

\subsection{Test with elliptical galaxies: fitting data}
 To assemble a sample for dwarf spheroidal (dSph) galaxies, we choose 43 dSphs from the sample of all dwarf galaxies in and around the Local Group, as presented in~\cite{McConnachie2012AJ_Dwarf}. The sample we have selected includes information such as the effective radius $R_e$, velocity dispersion $\sigma^*$ and stellar mass $M_*$. This information is required when attempting to determine $\gamma^*$ and $\gamma_0$ in accordance with Equation (\ref{eq:Plummer Jeans equation}). We denote this sample as sample dSphs.
 
 Before proceeding further, it is essential to modify the effective radius $R_e$ for future use in conformal gravity (CG). In actual observations, the effective radius is measured in terms of angle $\theta_e=R_e/D_\text{A}(0,z)$, where $D_\text{A}(0,z)$ is the angular diameter distance to an object at redshit $z$. This angular diameter distance is derived in CG, its general formula is given in Equation (\ref{eq:angular diameter distance}). Given that $D_\text{A}(0,z)$ varies across different theory of gravity, if the data is presented in the framework of general relativity (GR), we should modify the value of $R^G_e$ according to
 \begin{equation}\label{eq:Re CG}
    R_e = R^G_e D_\text{A}(0,z) / D_\text{A}^G(0,z),
\end{equation}  
 where $R^G_e$ and $D_\text{A}^G(0,z)$ are the values evaluated in GR.
 
We employ the least square method to evaluate $\gamma^*$ and $\gamma_0$. From Equation (\ref{eq:Plummer Jeans equation}), let $y_{\text{obs}} = -5\sigma_\ast^2(r/R_e)^2r/(1+r^2/R_e^2) + V'_{\beta}(r)$ \, and \, 
 $x(\gamma^\ast,\gamma_0) = - V'_{\gamma}(r) -  V'_{\gamma_0}$, the $\chi^2$ is defined by
 \begin{equation}\label{eq:chi21}
    \chi^2 = \Sigma \left[\frac{y_{\text{obs}} - x(\gamma^\ast,\gamma_0)}{\sigma_{\text{obs}}}\right]^2.
\end{equation}
In actual calculations, we choose $\sigma_{\text{obs}}=1$. Since both sides of Equation (\ref{eq:Plummer Jeans equation}) are functions of radius $r$, we evaluate $\gamma*$ and $\gamma_0$ at $r=R_e$ for each galaxy. The optimized fitted values are as follows: $\gamma^\ast_{\text{dSph}} = 1.22 \times 10^{-35}m^{-1}$, $\gamma_{0,{\text{dSph}}} = 5.27 \times 10^{-28} \, m^{-1}$. By comparing the results obtained from fitting dwarf spheroidal galaxies with those from fitting spiral galaxies, as presented in Equation Equation (\ref{eq:value of gammas}), we find that $\gamma^*$ is four orders of magnitude larger, while $\gamma_0$ is of the same order. The universal value of $\gamma_0$  obtained from each dwarf spheroidal galaxy (dSph) is anticipated. This is because it stems from the cosmological effect on the local system and, consequently, is independent of any specific local gravitational system. However, the fact that the fitted value of $\gamma^\ast_{\text{dSph}}$  is much larger than that obtained from spiral galaxies implies that if the latter value is correct, it is insufficient to explain the dynamics of dSphs. In other words, when it comes to dSphs, a certain amount of dark matter must be introduced.

It should be noted that, as can be seen from the fitting results of the rotation curves of spiral galaxies in reference \citep{mannheim_fitting_2012}, when the stellar mass $M^\ast < 10^{11} (M_\odot)$, the $\gamma_0$ term dominates the linear potential. As $M_*$ increases, the $\gamma_*$ term gradually becomes dominant in the linear potential.  In our sample of dSphs, the stellar mass of all galaxies satisfies the condition $M^\ast < 10^{11} (M_\odot)$.  Therefore, the linear potential should be dominated by $\gamma_0$. On the other hand, in the conventional dark matter model, it is widely acknowledged that dark matter dominates the potential of dwarf galaxies, and this dominant tendency weakens as the stellar mass increases. In Conformal Gravity (CG), the Newtonian potential generated by dark matter is replaced by two linear potentials (i.e., $\gamma_0$ and $\gamma_*$ potentials). As shown in reference \citep{mannheim_fitting_2012}, for dwarf galaxies, there is a trend in CG that is similar to that in the dark matter model.  According to CG, both $\gamma_0$ and  $\gamma_*$, of course, should be universal constants.  However, as we will demonstrate, for dwarf spheroidal galaxies (dSphs), $\gamma^*$ is not a constant. Instead, it decreases with an increase in the stellar mass $M_*$. This tendency resembles the one in dark matter model but violate the basic assumptions of CG. To achieve this, we fix $\gamma_0$ to be a smaller value of $3.97 \times 10^{-29} \, m^{-1}$ (as opposed to the optimized value of $\gamma_{0,{\text{dSph}}} = 5.27 \times 10^{-28} \, m^{-1}$), and keep $\gamma^*$ as a free parameter to be determined. This fixed value of $\gamma_0$ is obtained by setting $\gamma^*=0$ for all galaxies and fitting the value of $\gamma_0$ according to the Jeans equation (\ref{eq:Plummer Jeans equation}), then finding the smallest one. Such a fixed value of $\gamma_0$ would ensure that the fitted value of $\gamma^*$ cannot be negative. The rationale behind choosing to fix $\gamma_0$ instead of $\gamma^*$ is as follows. In Conformal Gravity (CG), $\gamma^*$ represents the linear potential stemming from the local luminous mass. It can imitate the distribution of Dark Matter (DM) in Newtonian gravity within any local gravitational system. This enables us to draw a comparison between the DM distribution in Newtonian gravity and the linear potential generated by the luminous matter in CG. Conversely, $\gamma_0$ measures the cosmological impact on local systems, and this impact remains independent of any particular local system. We would like to emphasize that, as is evident from Equation (\ref{eq:total potential}), for a given galaxy, the combination of $\gamma_0$ and $N^*\gamma^*$ must remain a constant. Thus, a decrease in the value of $\gamma_0$ necessarily implies an increase in the value of $\gamma^*$. Nonetheless, fixing $\gamma_0$ at its smallest value will not impede our ability to draw a correct conclusion regarding the correlation between $M_*$ and $\gamma^*$. Meanwhile, we are aware that the actual optimal value of $\gamma_0$ for dSphs is $5.27 \times 10^{-28} \, m^{-1}$.

Subsequently, we will set $\gamma_0$ to $3.97 \times 10^{-29} \, m^{-1}$ and calculate $\gamma^*$ and $M_*$ at $r=R_e$  using the Jeans equations. We define $\chi^2$ 
\begin{equation}\label{eq:chi squre for M and gamma*}
\chi^2 = \displaystyle\sum_{i}^{\text{N}} \left( \log_{10} \gamma^\ast_i - a \log_{10} (M^*_i/M_\odot) - b \right)^2
\end{equation}
to obtain the optimal parameter $a$ and $b$ using least square method. By doing so, we can establish the expected correlation between the stellar mass $M_*(R_e)$ and $\gamma^*(R_e)$. For our selected sample of 43 dSphs, by applying Equation (\ref{eq:Plummer Jeans equation}) to Equation (\ref{eq:chi squre for M and gamma*}) we find an empirical formula 
\begin{equation}\label{eq:empirical formula between M and gamma* for dSphs }
\gamma^\ast_{\text{dSph}} = 2.75 \times 10^{-28} (M_*/M_\odot)^{-0.963} \, m^{-1}.
\end{equation}

\begin{figure}
    \centering
   \includegraphics[width=8cm, angle=0]{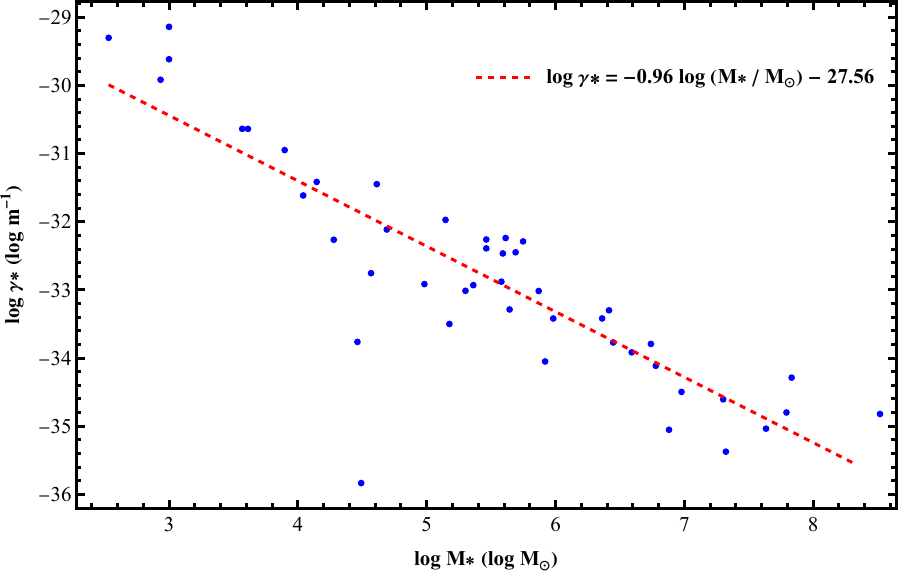}
    \caption{The correlation between the stellar mass  $M_*$ and $\gamma^\ast_{\text{dSph}}$. The results are obtained based on our selected sample for dSphs. We have set $\gamma_0$ to $3.97 \times 10^{-29} \, m^{-1}$ and calculate $\gamma^*$ at $r=R_e$  using the Jeans equation (\ref{eq:Plummer Jeans equation}.)} 
    \label{fig:PlotGammaStarPlummerDSph1}
\end{figure}

The results are shown in Fig.~\ref{fig:PlotGammaStarPlummerDSph1}. \, Evidently, $\gamma^\ast_{\text{dSph}}$ is not a constant as predicted by Conformal Gravity (CG). Instead, it decreases as the stellar mass $M_*$ increases. 

\begin{figure}
    \centering
   \includegraphics[width=8cm, angle=0]{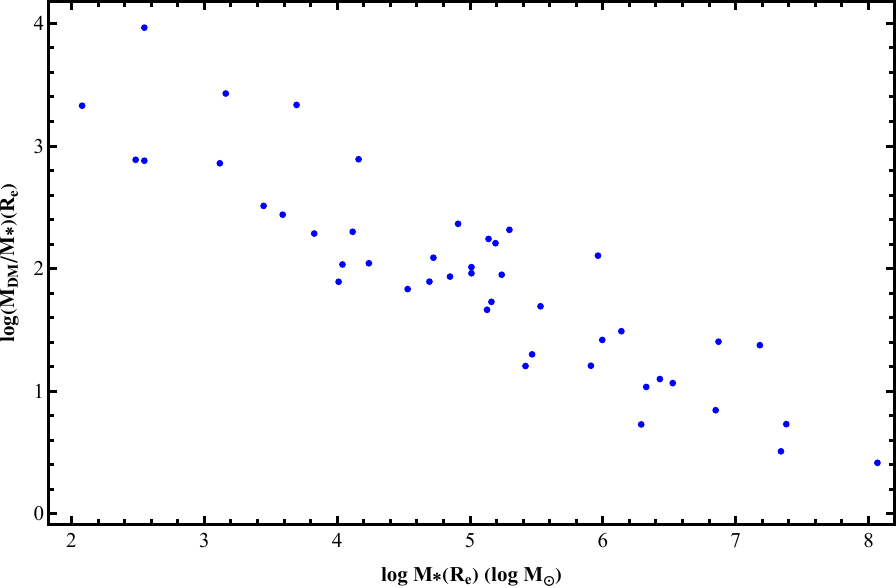}
    \caption{The correlation between $M_\text{DM}$ and $M_*$ for the sample of dSphs.} 
    \label{PlotGammaStarPlummerDSph2}
\end{figure}

It would be intriguing to explore the correlation between the dark matter mass $M_\text{dyn}$ and the stellar mass $M_*$ in Newtonian gravity and to check whether this correlation resembles that between $\gamma^*$ and $M_*$ in CG. To accomplish this, we aplly the Jeans equation (\ref{eq:MJSPul}) for dSphs. For simplicity's sake, we calculate the total mass $M_\text{dyn}(R_e)$ and stellar mass $M_*(R_e)$ within $r=R_e$. Thus, the dark matter mass within $R_e$ is $M_\text{DM}(R_e)=M_\text{dyn}(R_e)-M_*(R_e)$.  The correlation between $M_\text{DM}(R_e)$ and $M_*(R_e)$ is depicted in Fig.~\ref{PlotGammaStarPlummerDSph2}. Indeed, this figure reveals that the dark matter mass $M_\text{DM}(R_e)$ decreases as the stellar mass  $M_*(R_e)$ increases, following exactly the same patern as that of $\gamma^\ast_{\text{dSph}}$ and $M_*$ as shown in Fig.~\ref{fig:PlotGammaStarPlummerDSph1}. 

For ease of reference, we list in Table~\ref{tab:PlummerDSphtxt}  the following parameters for each galaxy in our selected sample, which is based on the sample in reference \cite{McConnachie2012AJ_Dwarf}: the total stellar mass $M_*$, effective radius $R_e$, the fitted value of $\gamma^\ast_{\text{dSph}}$, the dynamical mass $M_\text{dyn}(R_e)$ within $R_e$ and the dark matter-stellar mass ratio $M_\text{dyn}(R_e)/M_*(R_e)$ within $R_e$.

We now shift our focus to the investigation of $\gamma^*$ for bright elliptical galaxies. We will apply our method used for Dwarf Spheroidal Galaxies (dSphs) to two samples of bright elliptical galaxies. 

\begin{figure}
    \centering
   \includegraphics[width=8cm, angle=0]{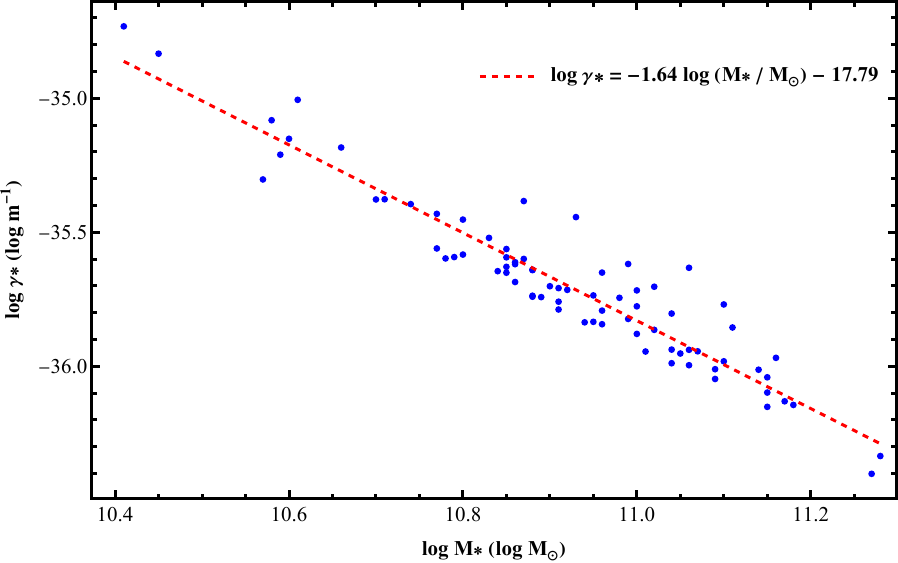}
    \caption{The correlation between $\gamma^\ast_{\text{SDSS}}$ and $M_*$ based on the sample of SDSS DR 10~\citep{Saulder2015AA}.} 
    \label{PlotGammaStarSDSSDR1}
\end{figure}

The first sample is composed of 76 compact, high velocity-dispersion, early-type galaxies from the Sloan Digital Sky Survey (SDSS) with $0.05<z<0.2$. We denote this sample as SDSS DR 10. This sample was established in reference \citep{Saulder2015AA} by employing de Vaucouleurs model (Sersic profile with $n=4$). Therefore, for bight galaxies, we utilise Jeans equation (\ref{eq:Jeans equation for Sersic profile}) to calculate $\gamma_\text{SDSS}^*$ at $r=R_e$. As proposed in reference~\citep{Saulder2015AA}, in this scenario, $n=4, \, bn=7.66925$ and $p=0.854938$. Meanwhile, the effective radius $R_e$ is adjusted in accordance with Equation (\ref{eq:Re CG}). Because of the large velocity dispersion, an correction is required and we take advantage of the work of~\cite{Shu2014THESL} and ~\cite{Cappellari2005TheSP} to use
\begin{equation}\label{eq:SDSSAg}
    \sigma_e =  \sigma_{\text{SDSS}} {(\frac{1.5''}{\theta_e})}^{0.05}
\end{equation}
as the corrected velocity dispersion at $R_e$, where $1.5''$ is the angular radius of the SDSS fiber, and $\theta_e$ is the effective radius. The results indicate that the correlation between $\gamma_\text{SDSS}^*$ and $M_*$ can be described by a fitted formula $\gamma^\ast_{\text{SDSS}} = 1.62 \times 10^{-18} (M_*/M_\odot)^{-1.64} \, m^{-1}$. Similar to the case of dwarf spheroidal galaxies, $\gamma_\text{SDSS}^*$ is not a constant; instead, it decreases as $M_*$ increases, as presented in Fig.~\ref{PlotGammaStarSDSSDR1}. 

\begin{figure}
    \centering
   \includegraphics[width=8cm, angle=0]{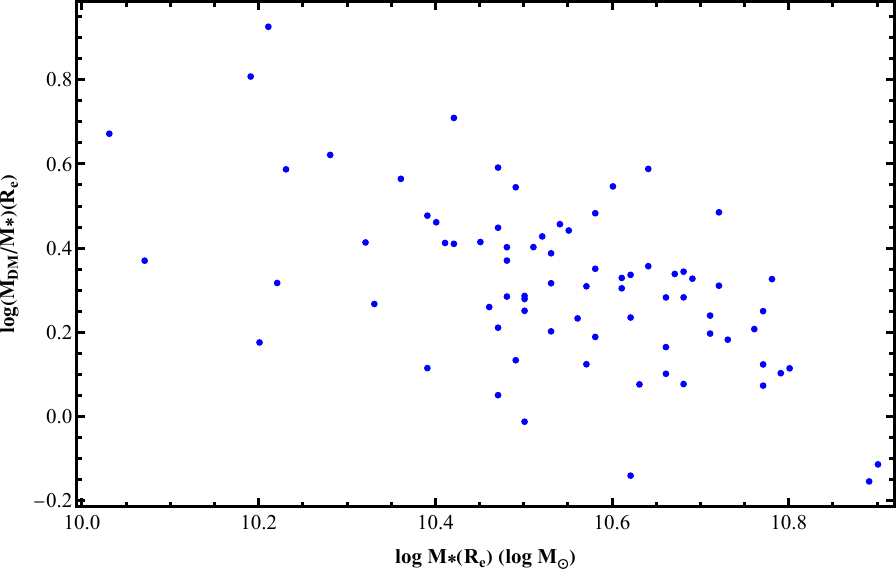}
    \caption{The correlation between $M_\text{DM}$ and $M_*$ derived from the sample of SDSS DR 10~\citep{Saulder2015AA}.} 
    \label{PlotGammaStarSDSSDR2}
\end{figure}

Meanwhile, the mass of dark matter $M_\text{DM}(R_e)$ is calculated according to Equation (\ref{eq:MJS}). The correlation between $M_\text{DM}(R_e)$ and $M_*$ is presented in Fig.~\ref{PlotGammaStarSDSSDR2}. As depicted, this correlation is weak. However, in a certain sense, it is still similar to that between $\gamma^\ast_{\text{SDSS}}$ and $M_*$. 

The relevant original and derived  parameters from sample SDSS10 are listed in Table~\ref{SDSSDR10txt}.

\begin{figure}
    \centering
   \includegraphics[width=8cm, angle=0]{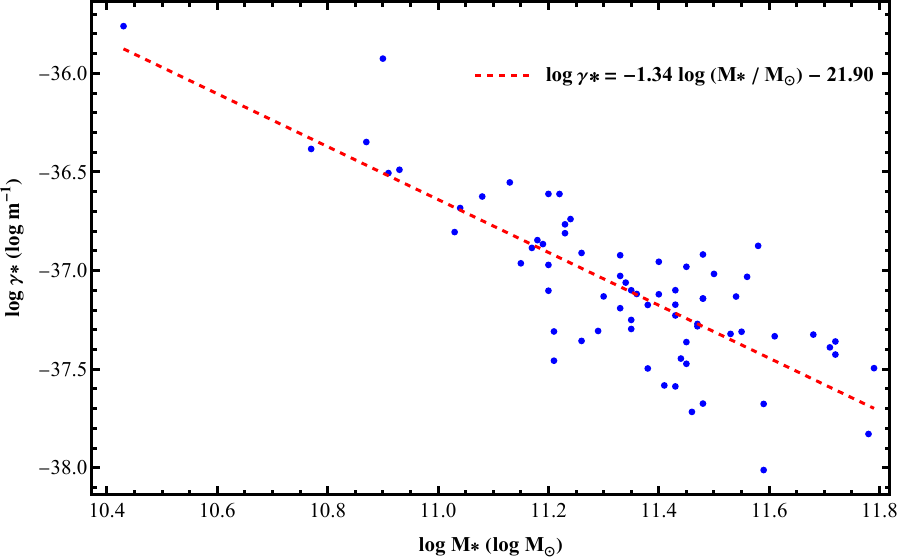}
    \caption{the correlation between $\gamma^\ast_{\text{SLACS}}(R_e)$ and $M_*(R_e)$ based on sample SLACS. } 
    \label{PlotGammaStarSLACS1}
\end{figure}

\begin{figure}
    \centering
   \includegraphics[width=8cm, angle=0]{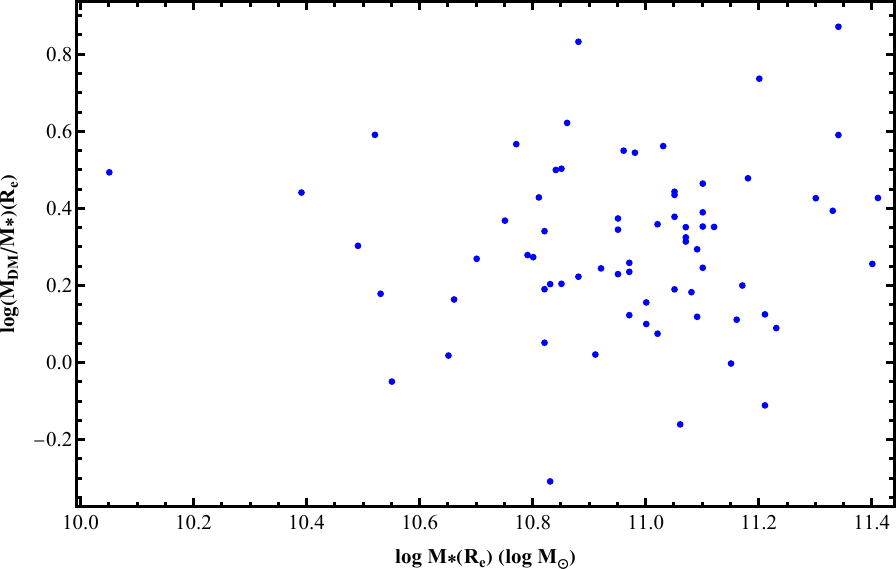}
    \caption{The correlation between $M_\text{DM}$ and $M_*$ derived from the sample SLACS.} 
    \label{PlotGammaStarSLACS2}
\end{figure}

To extract out more information about $\gamma^*$ from bright galaxies,  we use a new sample based on the data set of the Sloan Lens ACS (SLACS) Survey \citep{2009ApJ...705.1099A} to carry out the same procedure as we did for the sample SDSS DR 10. This data set was originally used for gravitational lensing analysis, but it provides us with more information that we need to study the properties of $\gamma^*$. We denote this data set as sample SLACS. Compared with the sample SDSS DR 10, the galaxies in sample SLACS are brighter and have a larger effective radius. The correlation between $\gamma^\ast_{\text{SLACS}}$ and $M_*$ based on sample SLACS is presented in Fig.~\ref{PlotGammaStarSLACS1}. As shown, the correlation between  $\gamma^\ast_{\text{SLACS}}$ and $M_*$ is weaker than that for $\gamma^\ast_{\text{SDSS}}$ and that for $\gamma^\ast_{\text{dSph}}$. The correlation between $M_\text{DM}(R_e)$ and $M_*$ for sample SLACS is also shown in Fig.~\ref{PlotGammaStarSLACS2}. As is evident, the correlation is much weaker than that for sample dSphs and that for sample SDSS DR 10.  The parameters for sample SLACS are also presented in Table~\ref{SLACStxt}.
 
\section{Conclusions and discussions}\label{sec:conclusions}
An exact Conformal Gravity (CG) analog of the Schwarzschild exterior solution was found to predict a linear potential  $V_{\gamma}=\frac{1}{2}\gamma c^2 r$  besides the conventional Newtonian potential $V_{\beta}=-\frac{\beta c^2}{r}$~\cite{1989ApJ...342..635M}. It was also found that there exists a universal, galaxy-independent linear potential, $V_{\gamma_0}=\frac{1}{2}\gamma_0 c^2r$, due to the rest of matter in the universe on any local galaxies~\citep{Mannheim_1997}. The parameter $\gamma=(M/M_{\sun})\gamma^*$, where $M$ is the mass of luminous matter that generates the corresponding linear potential $V_{\gamma}$, and $\gamma^*$ is the value of $\gamma$ if $M=M_{\sun}$. Hence, the values of $\gamma_0$ and $\gamma^*$ should be universal constants independent of galaxies. These predictions of CG can be verified through galaxy observations. To date in the literature, the tests have been successfully conducted only via the observations of spiral galaxies, specifically using the rotation curves data. The rich data of this kind uniformly gives $\gamma^\ast=5.42 \times 10^{-39} m^{-1}$ and $\gamma_0=3.06 \times 10^{-28} m^{-1}$. 

In contrast, in this paper, we aim to test CG by utilizing the velocity dispersion data from the observations of elliptical galaxies. It is well known that within elliptical galaxies, an extra gravitational force is required to balance the observed velocity dispersion. The Jeans equation is a useful tool for describing the relationship between the velocity dispersion and the gravitational potential. The Jeans equation was originally developed in Newtonian theory. In this context, dark matter is introduced to account for the extra potential. To test Conformal Gravity (CG), we extend the Jeans equation by simply replacing the Newtonian potential with the potential predicted by CG. In fact, when people apply CG to spiral galaxies, they follow the same approach. That is, they replace the Newtonian potential with the potential predicted by CG to explain the observed rotation curves.  

We first select a sample, sample dSphs, consisting of 43 dwarf spheroidal galaxies (dSphs) based on the reference~\cite{McConnachie2012AJ_Dwarf}. We found that the value of $\gamma_0=5.27 \times 10^{-28} \, m^{-1}$ derived from the observations of the elliptical galaxies has the same order of that derived from the observations of spiral galaxies. This result is not surprising, since the $\gamma_0$ term in linear potentials originates from the cosmological effect on any local gravitational systems, and thus should be independent of local systems. However, our sample dSphs gives the optimum value of $\gamma^\ast_{\text{dSph}} = 1.22 \times 10^{-35}m^{-1}$, which is about four orders of magnitude larger than that fitted by spiral galaxies ($\sim 10^{-39} \, m^{-1}$). It suggests that the linear potential of luminous matter estimated from spiral galaxies is negligible when applied to elliptical galaxies. This inconsistent result between elliptical and spiral galaxies may indicate that Conformal Gravity (CG) fails as an alternative to the dark matter model, at least for elliptical galaxies.

Furthermore, as depicted in Fig.~\ref{fig:PlotGammaStarPlummerDSph1}, we discover a strong correlation between $\gamma^\ast_{\text{dSph}}(R_e)$ and the stellar mass $M_*(R_e)$ for dwarf spheroidal galaxies. This is accomplished by fixing $\gamma_0$ and treating $\gamma^*$ as a free parameter. As evident from Fig.\ref{fig:PlotGammaStarPlummerDSph1}, $\gamma^*$ decreases as $M^*$ increases. Interestingly enough, this situation is analogous to that in Newtonian gravity, where dark matter is introduced to provide the necessary extra potential. In Newtonian gravity, it is a widely - accepted notion that the brighter the galaxy, the less dark matter is required. In fact, we applied Newtonian gravity to the same sample and calculated the dark matter mass and the luminous stellar mass within the effective radius for each galaxy. as shown in Fig.~\ref{PlotGammaStarPlummerDSph2}, we found a correlation between dark matter mass $M_\text{DM}$ and stellar mass $M_*$ that is similar to the correlation between $\gamma^\ast_{\text{dSph}}$ and stellar mass $M_*$. These results imply that, to explain the observations of dwarf spheroidal galaxies, $\gamma^\ast_{\text{dSph}}$ cannot be a constant. Instead, it behaves more like the amount of dark matter, which can vary with the amount of stellar matter. Regrettably, the varying value of $\gamma^*$ violates the fundamental prediction of CG, which requires $\gamma^*$ to be a universal constant. 

Dwarf spheroidal galaxies (dSphs) are dominated by an extra gravitational potential. It would be interesting to explore the correlations we discovered in dSphs using the data sets of bright elliptical galaxies. To this end, we selected two samples: sample SDSS DR 10 and sample SLACS. The galaxies in sample SLACS are brighter than those in sample SDSS DR 10. We carried out the same procedure as we did for dSphs. For sample SDSS DR 10, we found that the correlation between $\gamma_\text{SDSS}^*$ and $M_*$ (as shown in Fig.~\ref{PlotGammaStarSDSSDR1})  and the correlation between $M_\text{DM}(R_e)$ and $M_*$ (as shown in Fig.~\ref{PlotGammaStarSDSSDR2})are weaker than the corresponding correlations for sample dSphs. We further found that the correlations for sample SLACS (as shown in Fig.~\ref{PlotGammaStarSLACS1} and Fig.~\ref{PlotGammaStarSLACS2}) are even weaker than those for sample SDSS DR 10. This indicates that when less extra potential is needed, the correlations are statistically more scattered, as expected. For ease of reference, we list all the parameters of each sample in the corresponding table. 

As shown in Equations (\ref{eq:potential of the sun}) and (\ref{eq:potential of any mass M}), $\gamma^*$ characterises the linear potential of a unit mass, and should therefore be a universal constant. However, the observed correlation between $\gamma^*$ and the stellar mass $M_*$ in galaxies closely resembles the correlation between $M_\text{DM}$ and stellar mass $M_*$. This suggests that elliptical galaxies are better described by Newtonian theory (requiring dark matter) than by conformal gravity. Of course, this does not necessarily mean that Newtonian gravity is correct unless dark matter particles are directly detected in experiments. Alternatively, the $\gamma^*–M_*$ correlation could imply an additional scale-dependent quantum potential in large scale structures, as proposed by~\citep{Chen_2022,universe10080333}.

 \begingroup
\footnotesize
\begin{longtable}{cccccccc}
  \caption[]{Parameters of Each Galaxy in Our Selected Sample for dSphs. From Column 1 to column 7: Galaxy Name, Total Stellar Mass ($M_*$), Effective Radius ($R_e$), Velocity Dispersion ($\sigma_e$), Fitted Value of $\gamma^\ast_{\text{dSph}}$, Dynamical Mass ($M_\text{dyn}(R_e)$) within $R_e$ and Dark Matter-Stellar Mass Ratio $\left(M_\text{dyn}(R_e)/M_*(R_e)\right)$ within $R_e$. \label{tab:PlummerDSphtxt}}\\
  \hline\noalign{\smallskip}
\hline\noalign{\smallskip}
  Galaxy     & $\log_{10}(M_\ast)$    & $R_{\text{e}}$   & $\sigma_{\text{e}}$    & $\log_{10}\gamma^\ast_{\text{dSph}} $    & $\log_{10}(M_{\text{dyn}})$ & $\log_{10} A(<R_{\text{e}})$ \\
             & [$\log_{10}(M_\odot)$] & [pc]             & [km s$^{-1}$]          & [$\log_{10}(m^{-1})$]                    & [$\log_{10}(M_\odot)$]      & A = $M_{\text{DM}}/M_\ast$   \\
  \hline\noalign{\smallskip}
  \endfirsthead
  \multicolumn{7}{c}{\tablename\ \thetable\ -- \textit{continued}}\\
  \hline\noalign{\smallskip}
  Galaxy     & $\log_{10}(M_\ast)$    & $R_{\text{e}}$   & $\sigma_{\text{e}}$    & $\log_{10}\gamma^\ast_{\text{dSph}} $    & $\log_{10}(M_{\text{dyn}})$ & $\log_{10} A(<R_{\text{e}})$ \\
             & [$\log_{10}(M_\odot)$] & [pc]             & [km s$^{-1}$]          & [$\log_{10}(m^{-1})$]                    & [$\log_{10}(M_\odot)$]      & A = $M_{\text{DM}}/M_\ast$   \\
  \hline\noalign{\smallskip}
  \endhead
  \hline\noalign{\smallskip}
  \endfoot
  \hline\noalign{\smallskip}
  \endlastfoot
  Sagittarius dSph  & 7.32  & 2587 & 11.4   & -35.37    & 8.29  & 1.40     \\  
  Segue (I)         & 2.53  & 29   & 3.9    & -29.30    & 5.41  & 3.33     \\  
  Ursa Major II     & 3.61  & 149  & 6.7    & -30.64    & 6.59  & 3.43     \\  
  Bootes II         & 3.00  & 51   & 10.5   & -29.14    & 6.51  & 3.97     \\  
  Segue II          & 2.93  & 35   & 3.4    & -29.92    & 5.37  & 2.89     \\  
  Willman 1         & 3.00  & 25   & 4.3    & -29.62    & 5.43  & 2.88     \\  
  Coma Berenices    & 3.57  & 77   & 4.6    & -30.64    & 5.98  & 2.86     \\  
  Bootes (I)        & 4.46  & 242  & 2.4    & -33.76    & 5.91  & 1.89     \\  
  Draco             & 5.46  & 221  & 9.1    & -32.39    & 7.03  & 2.01     \\  
  Ursa Minor        & 5.46  & 181  & 9.5    & -32.26    & 6.98  & 1.96     \\  
  Sculptor          & 6.36  & 283  & 9.2    & -33.42    & 7.14  & 1.21     \\  
  Sextans (I)       & 5.64  & 695  & 7.9    & -33.29    & 7.40  & 2.21     \\  
  Ursa Major (I)    & 4.15  & 319  & 7.6    & -31.42    & 7.03  & 3.33     \\  
  Carina            & 5.58  & 250  & 6.6    & -32.88    & 6.80  & 1.66     \\  
  Hercules          & 4.57  & 330  & 3.7    & -32.75    & 6.42  & 2.30     \\  
  Fornax            & 7.30  & 710  & 11.7   & -34.61    & 7.75  & 0.84     \\  
  Leo IV            & 4.28  & 206  & 3.3    & -32.26    & 6.12  & 2.29     \\  
  Canes Venatici II & 3.90  & 74   & 4.6    & -30.95    & 5.96  & 2.51     \\  
  Leo V             & 4.04  & 135  & 3.7    & -31.61    & 6.03  & 2.44     \\  
  CanesVenatici (I) & 5.36  & 564  & 7.6    & -32.93    & 7.28  & 2.37     \\  
  Leo II            & 5.87  & 176  & 6.6    & -33.02    & 6.65  & 1.21     \\  
  Leo I             & 6.74  & 251  & 9.2    & -33.79    & 7.09  & 0.73     \\  
  Andromeda IX      & 5.18  & 557  & 4.5    & -33.50    & 6.82  & 2.09     \\  
  NGC 205           & 8.52  & 590  & 35.0   & -34.82    & 8.62  & 0.42     \\  
  Andromeda I       & 6.59  & 672  & 10.6   & -33.92    & 7.64  & 1.49     \\  
  Andromeda III     & 5.92  & 479  & 4.7    & -34.05    & 6.79  & 1.30     \\  
  Andromeda XI      & 4.69  & 157  & 4.6    & -32.11    & 6.29  & 2.04     \\  
  Andromeda X       & 4.98  & 265  & 3.9    & -32.91    & 6.37  & 1.83     \\  
  Andromeda XII     & 4.49  & 304  & 2.6    & -35.83    & 6.08  & 2.03     \\  
  NGC 147           & 7.79  & 623  & 16.0   & -34.80    & 7.97  & 0.51     \\  
  Andromeda XIV     & 5.30  & 363  & 5.4    & -33.01    & 6.79  & 1.93     \\  
  Andromeda XV      & 5.69  & 222  & 11.0   & -32.45    & 7.19  & 1.95     \\  
  Andromeda XIII    & 4.61  & 207  & 9.7    & -31.45    & 7.05  & 2.89     \\  
  Andromeda II      & 6.88  & 1176 & 7.3    & -35.05    & 7.56  & 1.10     \\  
  NGC 185           & 7.83  & 458  & 24.0   & -34.29    & 8.19  & 0.73     \\  
  Andromeda VII     & 6.98  & 776  & 9.7    & -34.50    & 7.63  & 1.07     \\  
  LGS 3             & 5.98  & 470  & 7.9    & -33.42    & 7.23  & 1.69     \\  
  Andromeda XVI     & 5.61  & 136  & 10.0   & -32.24    & 6.90  & 1.73     \\  
  Cetus             & 6.41  & 703  & 17.0   & -33.30    & 8.07  & 2.11     \\  
  Leo T             & 5.15  & 120  & 7.5    & -31.97    & 6.59  & 1.89     \\  
  *WLM*             & 7.63  & 2111 & 17.5   & -35.04    & 8.57  & 1.38     \\  
  *Leo A*           & 6.78  & 499  & 9.3    & -34.11    & 7.40  & 1.03     \\  
  Tucana            & 5.75  & 284  & 15.8   & -32.29    & 7.61  & 2.32     \\  
  average           & 7.12  & 436  & 8.94   & -30.43    & 7.62  & 2.70     \\ 
\end{longtable}
\endgroup

\begingroup
\footnotesize
\begin{longtable}{cccccccc}
  \caption[]{The relevant original and derived parameters from sample SDSS DR 10. From column 1 to column 8: Object IDs, redshift $z$, stellar mass $M_*(R_e)$, effective radius $R_e$, velocity dispersion $\sigma_e$, $\gamma^*_{\text{SDSS}}$, total mass $M_\text{dyn}(R_e)$, and the mass ratio of dark matter to luminous matter. \label{SDSSDR10txt}}\\
  \hline\noalign{\smallskip}
  \hline\noalign{\smallskip}
  Galaxy     & z & $\log_{10}(M_\ast)$   & $R_{\text{e}}$  & $\sigma_{\text{e}}$    & $\log_{10}\gamma^\ast_{\text{SDSS}} $    & $\log_{10}(M_{\text{dyn}})$ & $\log_{10} A(<R_{\text{e}})$ \\
             &   & [$\log_{10}(M_\odot)$]& [kpc]           & [km s$^{-1}$]          & [$\log_{10}(m^{-1})$]                    & [$\log_{10}(M_\odot)$]      & A = $M_{\text{DM}}/M_\ast$   \\
  \hline\noalign{\smallskip}
  \endfirsthead
  \multicolumn{8}{c}{\tablename\ \thetable\ -- \textit{continued}}\\
  \hline\noalign{\smallskip}
  Galaxy     & z & $\log_{10}(M_\ast)$   & $R_{\text{e}}$  & $\sigma_{\text{e}}$    & $\log_{10}\gamma^\ast_{\text{SDSS}} $    & $\log_{10}(M_{\text{dyn}})$ & $\log_{10} A(<R_{\text{e}})$ \\
             &   & [$\log_{10}(M_\odot)$]& [kpc]           & [km s$^{-1}$]          & [$\log_{10}(m^{-1})$]                    & [$\log_{10}(M_\odot)$]      & A = $M_{\text{DM}}/M_\ast$   \\
  \hline\noalign{\smallskip}
  \endhead
  \hline\noalign{\smallskip}
 
  \endfoot
  \hline
  \endlastfoot
  SDSS J154713.73$-$000831.8 & 0.1138 & 11.04 & 2.00 & 321.97 & $-$35.94      & 11.13  & 0.28         \\     
  SDSS J151741.75$-$004217.4 & 0.1166 & 11.15 & 2.17 & 342.40 & $-$36.04      & 11.21  & 0.25         \\     
  SDSS J082216.57$+$481519.0 & 0.1276 & 10.98 & 2.17 & 359.08 & $-$35.74      & 11.26  & 0.55         \\     
  SDSS J105603.78$+$015953.8 & 0.1153 & 10.77 & 1.62 & 306.96 & $-$35.56      & 10.99  & 0.48         \\     
  SDSS J214923.79$-$084030.5 & 0.1014 & 10.83 & 1.44 & 330.82 & $-$35.52      & 11.01  & 0.41         \\     
  SDSS J035212.98$-$055140.0 & 0.1137 & 10.95 & 1.32 & 319.33 & $-$35.74      & 10.94  & 0.12         \\     
  SDSS J003241.18$-$103958.0 & 0.1557 & 11.16 & 2.18 & 366.14 & $-$35.97      & 11.28  & 0.33         \\     
  SDSS J163138.81$+$461605.7 & 0.1321 & 11.00 & 1.02 & 330.89 & $-$35.78      & 10.86  & $-$0.14      \\       
  SDSS J170541.78$+$332840.3 & 0.1022 & 10.85 & 2.05 & 331.24 & $-$35.65      & 11.16  & 0.59         \\     
  SDSS J111052.92$+$664710.4 & 0.1362 & 10.99 & 1.42 & 366.75 & $-$35.62      & 11.09  & 0.30         \\     
  SDSS J143314.96$+$013019.1 & 0.1096 & 10.88 & 1.58 & 300.47 & $-$35.74      & 10.96  & 0.28         \\     
  SDSS J161348.81$+$410621.1 & 0.1381 & 11.01 & 1.59 & 302.34 & $-$35.95      & 10.97  & 0.08         \\     
  SDSS J012316.92$+$001743.9 & 0.0928 & 10.78 & 1.66 & 302.65 & $-$35.60      & 10.99  & 0.46         \\     
  SDSS J163318.88$+$470738.8 & 0.1229 & 10.66 & 1.24 & 351.87 & $-$35.18      & 11.00  & 0.62         \\     
  SDSS J125411.36$+$504901.3 & 0.1209 & 11.00 & 1.65 & 352.84 & $-$35.72      & 11.12  & 0.34         \\     
  SDSS J000431.74$+$160418.7 & 0.1144 & 10.85 & 1.03 & 307.55 & $-$35.59      & 10.80  & 0.05         \\     
  SDSS J154220.18$+$044559.9 & 0.1105 & 10.95 & 1.83 & 309.66 & $-$35.83      & 11.05  & 0.31         \\     
  SDSS J004130.42$-$091406.6 & 0.0538 & 11.04 & 1.86 & 307.47 & $-$35.99      & 11.05  & 0.16         \\     
  SDSS J220706.06$+$120245.2 & 0.1607 & 11.15 & 2.13 & 316.38 & $-$36.15      & 11.14  & 0.12         \\     
  SDSS J233639.48$+$154919.9 & 0.1179 & 10.88 & 1.60 & 300.21 & $-$35.74      & 10.97  & 0.29         \\     
  SDSS J162225.18$+$444708.3 & 0.0716 & 11.02 & 2.01 & 333.39 & $-$35.86      & 11.16  & 0.36         \\     
  SDSS J081512.33$+$384045.4 & 0.1259 & 11.14 & 2.00 & 341.86 & $-$36.01      & 11.18  & 0.21         \\     
  SDSS J143133.11$+$085520.9 & 0.1108 & 11.02 & 2.17 & 390.45 & $-$35.70      & 11.33  & 0.59         \\     
  SDSS J165802.87$+$415016.0 & 0.0375 & 10.60 & 0.85 & 307.34 & $-$35.15      & 10.71  & 0.32         \\     
  SDSS J095532.65$+$042219.7 & 0.0937 & 11.11 & 1.62 & 360.50 & $-$35.86      & 11.13  & 0.18         \\     
  SDSS J121921.58$+$633208.8 & 0.1039 & 10.77 & 0.92 & 309.05 & $-$35.43      & 10.75  & 0.11         \\     
  SDSS J123045.21$+$514221.4 & 0.1517 & 11.06 & 1.58 & 321.43 & $-$35.94      & 11.02  & 0.08         \\     
  SDSS J224144.94$-$004840.7 & 0.1293 & 10.93 & 1.36 & 390.83 & $-$35.44      & 11.13  & 0.44         \\     
  SDSS J091318.85$+$080658.0 & 0.0934 & 10.85 & 1.29 & 305.54 & $-$35.63      & 10.89  & 0.21         \\     
  SDSS J102516.66$+$401855.2 & 0.0682 & 10.57 & 1.77 & 318.69 & $-$35.30      & 11.06  & 0.81         \\     
  SDSS J000224.65$+$003206.5 & 0.0784 & 10.41 & 0.78 & 348.80 & $-$34.73      & 10.79  & 0.67         \\     
  SDSS J104047.00$+$395551.8 & 0.1394 & 10.96 & 1.29 & 341.47 & $-$35.65      & 10.99  & 0.19         \\     
  SDSS J150508.55$+$300706.1 & 0.1450 & 11.15 & 1.89 & 325.21 & $-$36.10      & 11.11  & 0.07         \\     
  SDSS J120100.67$+$121303.0 & 0.1295 & 10.94 & 1.68 & 301.72 & $-$35.84      & 10.99  & 0.23         \\     
  SDSS J145233.30$+$223533.6 & 0.1551 & 11.18 & 2.09 & 328.62 & $-$36.14      & 11.16  & 0.11         \\     
  SDSS J155454.67$+$252808.7 & 0.1556 & 11.09 & 2.16 & 317.81 & $-$36.05      & 11.15  & 0.24         \\     
  SDSS J110705.69$+$131905.3 & 0.1188 & 10.96 & 2.13 & 334.83 & $-$35.79      & 11.19  & 0.48         \\     
  SDSS J103205.36$+$372808.1 & 0.1043 & 11.06 & 1.51 & 397.57 & $-$35.63      & 11.19  & 0.34         \\     
  SDSS J145217.61$+$222913.5 & 0.1165 & 10.87 & 0.89 & 356.93 & $-$35.38      & 10.86  & 0.13         \\     
  SDSS J095626.81$+$235750.9 & 0.1193 & 10.80 & 1.87 & 365.85 & $-$35.45      & 11.21  & 0.71         \\     
  SDSS J233528.04$+$010248.2 & 0.0827 & 11.09 & 1.98 & 322.72 & $-$36.01      & 11.12  & 0.20         \\     
  SDSS J155816.65$+$271412.2 & 0.0896 & 10.88 & 1.01 & 309.78 & $-$35.64      & 10.80  & $-$0.01      \\       
  SDSS J083437.14$+$241930.1 & 0.0705 & 10.84 & 1.39 & 301.57 & $-$35.65      & 10.91  & 0.26         \\     
  SDSS J080654.35$+$204544.4 & 0.1247 & 10.91 & 1.40 & 312.35 & $-$35.71      & 10.94  & 0.20         \\     
  SDSS J131759.74$+$433650.9 & 0.1140 & 10.71 & 1.03 & 303.27 & $-$35.38      & 10.79  & 0.27         \\     
  SDSS J143637.07$+$312339.4 & 0.0850 & 11.04 & 1.39 & 340.11 & $-$35.80      & 11.02  & 0.10         \\     
  SDSS J115449.46$+$262556.4 & 0.1108 & 10.90 & 1.80 & 324.32 & $-$35.70      & 11.09  & 0.43         \\     
  SDSS J080651.62$+$192759.1 & 0.1242 & 10.70 & 1.22 & 309.2  & $-$35.38      & 10.88  & 0.41         \\     
  SDSS J140009.03$+$355701.1 & 0.1494 & 11.05 & 2.17 & 326.47 & $-$35.95      & 11.17  & 0.34         \\     
  SDSS J161312.98$+$174828.7 & 0.0374 & 10.91 & 1.70 & 309.47 & $-$35.76      & 11.02  & 0.32         \\     
  SDSS J223218.80$-$002421.2 & 0.0865 & 10.74 & 1.48 & 335.21 & $-$35.40      & 11.03  & 0.56         \\     
  SDSS J125705.31$+$285852.9 & 0.0686 & 11.10 & 2.18 & 338.34 & $-$35.98      & 11.20  & 0.31         \\     
  SDSS J120711.64$+$235227.9 & 0.0775 & 10.85 & 1.57 & 333.58 & $-$35.56      & 11.05  & 0.45         \\     
  SDSS J150913.80$+$162559.7 & 0.1159 & 10.86 & 1.52 & 321.50 & $-$35.61      & 11.01  & 0.37         \\     
  SDSS J160050.21$+$291210.0 & 0.0913 & 10.92 & 1.89 & 331.94 & $-$35.72      & 11.13  & 0.46         \\     
  SDSS J083546.02$+$341230.6 & 0.1978 & 11.28 & 1.99 & 330.50 & $-$36.34      & 11.15  & $-$0.11      \\       
  SDSS J135909.74$+$275700.3 & 0.0811 & 10.58 & 0.66 & 305.44 & $-$35.08      & 10.60  & 0.18         \\     
  SDSS J122035.75$+$291759.2 & 0.0908 & 11.07 & 2.12 & 335.01 & $-$35.94      & 11.19  & 0.33         \\     
  SDSS J125709.13$+$204823.2 & 0.0868 & 10.88 & 1.37 & 315.98 & $-$35.64      & 10.95  & 0.25         \\     
  SDSS J152811.97$+$120750.4 & 0.1225 & 10.86 & 1.59 & 322.58 & $-$35.62      & 11.03  & 0.40         \\     
  SDSS J151153.14$+$141555.0 & 0.1221 & 10.86 & 1.50 & 302.67 & $-$35.69      & 10.95  & 0.28         \\     
  SDSS J104112.53$+$001342.4 & 0.1300 & 11.00 & 1.77 & 315.47 & $-$35.88      & 11.06  & 0.23         \\     
  SDSS J124454.80$+$361101.7 & 0.0877 & 10.45 & 0.62 & 313.91 & $-$34.83      & 10.60  & 0.37         \\     
  SDSS J150340.60$+$171411.9 & 0.1505 & 11.06 & 2.14 & 318.55 & $-$36.00      & 11.15  & 0.28         \\     
  SDSS J154717.95$+$331038.1 & 0.1265 & 10.80 & 1.54 & 307.94 & $-$35.58      & 10.97  & 0.41         \\     
  SDSS J162230.11$+$092349.1 & 0.2018 & 11.27 & 2.05 & 315.64 & $-$36.40      & 11.12  & $-$0.15      \\       
  SDSS J120514.18$+$482517.8 & 0.0648 & 10.89 & 1.84 & 311.31 & $-$35.74      & 11.06  & 0.40         \\     
  SDSS J120951.59$+$200312.6 & 0.1116 & 10.79 & 1.56 & 302.97 & $-$35.59      & 10.97  & 0.41         \\     
  SDSS J162325.00$+$280527.4 & 0.1233 & 10.96 & 1.94 & 314.34 & $-$35.84      & 11.09  & 0.35         \\     
  SDSS J123952.07$+$210910.4 & 0.1085 & 11.17 & 2.03 & 327.18 & $-$36.13      & 11.15  & 0.10         \\     
  SDSS J141601.11$+$355927.7 & 0.1271 & 10.91 & 1.90 & 308.90 & $-$35.79      & 11.07  & 0.39         \\     
  SDSS J105003.10$+$114908.3 & 0.0812 & 10.87 & 1.83 & 343.88 & $-$35.60      & 11.14  & 0.54         \\     
  SDSS J150212.87$+$143803.5 & 0.0697 & 10.61 & 0.97 & 363.91 & $-$35.01      & 10.92  & 0.59         \\     
  SDSS J150430.87$+$063936.5 & 0.1439 & 10.99 & 1.85 & 327.53 & $-$35.82      & 11.11  & 0.33         \\     
  SDSS J141943.22$+$491411.9 & 0.0260 & 10.59 & 1.82 & 362.53 & $-$35.21      & 11.19  & 0.93         \\     
  SDSS J121607.29$+$210821.6 & 0.1278 & 11.10 & 2.08 & 398.82 & $-$35.77      & 11.33  & 0.48         \\
  average                  & 0.1107 & 10.95 & 1.63 & 328.81 & $-$35.57      & 11.06  & 0.36         \\      
\end{longtable}
\endgroup 

\begingroup
\footnotesize
\begin{longtable}{cccccccc}
  \caption[]{The relevant original and derived parameters from sample SLACS. From column 1 to column 8: Object IDs, redshift $z$, stellar mass $M_*(R_e)$, effective radius $R_e$, velocity dispersion $\sigma_e$, $\gamma^*_{\text{SDSS}}$, total mass $M_\text{dyn}(R_e)$, and the mass ratio of dark matter to luminous matter. \label{SLACStxt}}\\
  \hline\noalign{\smallskip}
  \hline\noalign{\smallskip}
  Galaxy     & z & $\log_{10}(M_\ast)$   & $R_{\text{e}}$  & $\sigma_{\text{e}}$    & $\log_{10}\gamma^\ast_{\text{SLACS}} $    & $\log_{10}(M_{\text{dyn}})$ & $\log_{10} A(<R_{\text{e}})$ \\
             &   & [$\log_{10}(M_\odot)$]& [kpc]           & [km s$^{-1}$]          & [$\log_{10}(m^{-1})$]                     & [$\log_{10}(M_\odot)$]      & A = $M_{\text{DM}}/M_\ast$   \\
  \hline\noalign{\smallskip}
  \endfirsthead
  \multicolumn{8}{c}{\tablename\ \thetable\ -- \textit{continued}}\\
  \hline\noalign{\smallskip}
  Galaxy     & z & $\log_{10}(M_\ast)$   & $R_{\text{e}}$  & $\sigma_{\text{e}}$    & $\log_{10}\gamma^\ast_{\text{SLACS}} $    & $\log_{10}(M_{\text{dyn}})$ & $\log_{10} A(<R_{\text{e}})$ \\
             &   & [$\log_{10}(M_\odot)$]& [kpc]           & [km s$^{-1}$]          & [$\log_{10}(m^{-1})$]                     & [$\log_{10}(M_\odot)$]      & A = $M_{\text{DM}}/M_\ast$   \\
  \hline\noalign{\smallskip}
  \endhead
  \hline\noalign{\smallskip}
 
  \endfoot
  \hline
  \endlastfoot
  SDSSJ0008$-$0004 & 0.440 & 11.38 & 10.34   & 191.14 & $-$37.50        & 11.39    & 0.16               \\
  SDSSJ0029$-$0055 & 0.227 & 11.33 & 9.35    & 222.87 & $-$37.19        & 11.48    & 0.37               \\
  SDSSJ0037$-$0942 & 0.195 & 11.48 & 8.64    & 271.02 & $-$37.14        & 11.61    & 0.35               \\
  SDSSJ0044$+$0113 & 0.120 & 11.23 & 7.03    & 255.91 & $-$36.81        & 11.47    & 0.50               \\
  SDSSJ0157$-$0056 & 0.513 & 11.50 & 7.51    & 298.19 & $-$37.02        & 11.63    & 0.35               \\
  SDSSJ0216$-$0813 & 0.332 & 11.79 & 14.12   & 321.82 & $-$37.49        & 11.98    & 0.43               \\
  SDSSJ0252$+$0039 & 0.280 & 11.21 & 5.75    & 164.81 & $-$37.46        & 11.00    & $-$0.31              \\
  SDSSJ0330$-$0020 & 0.351 & 11.35 & 7.50    & 211.86 & $-$37.25        & 11.34    & 0.12               \\
  SDSSJ0728$+$3835 & 0.206 & 11.44 & 6.76    & 210.89 & $-$37.45        & 11.29    & $-$0.16              \\
  SDSSJ0737$+$3216 & 0.322 & 11.72 & 15.74   & 324.55 & $-$37.43        & 12.03    & 0.59               \\
  SDSSJ0819$+$4534 & 0.194 & 11.15 & 8.99    & 218.09 & $-$36.96        & 11.44    & 0.57               \\
  SDSSJ0822$+$2652 & 0.241 & 11.43 & 9.21    & 252.83 & $-$37.17        & 11.58    & 0.38               \\
  SDSSJ0841$+$3824 & 0.116 & 11.41 & 18.22   & 206.10 & $-$37.58        & 11.70    & 0.56               \\
  SDSSJ0903$+$4116 & 0.430 & 11.59 & 12.28   & 218.82 & $-$37.68        & 11.58    & 0.12               \\
  SDSSJ0912$+$0029 & 0.164 & 11.71 & 12.05   & 309.31 & $-$37.39        & 11.87    & 0.39               \\
  SDSSJ0935$-$0003 & 0.347 & 11.72 & 20.17   & 376.49 & $-$37.36        & 12.27    & 0.87               \\
  SDSSJ0936$+$0913 & 0.190 & 11.43 & 7.90    & 236.87 & $-$37.23        & 11.46    & 0.19               \\
  SDSSJ0946$+$1006 & 0.222 & 11.34 & 9.87    & 255.06 & $-$37.06        & 11.62    & 0.55               \\
  SDSSJ0955$+$0101 & 0.111 & 10.77 & 4.01    & 189.35 & $-$36.38        & 10.97    & 0.44               \\
  SDSSJ0956$+$5100 & 0.241 & 11.56 & 8.79    & 326.80 & $-$37.03        & 11.78    & 0.48               \\
  SDSSJ0959$+$4416 & 0.237 & 11.47 & 7.64    & 240.28 & $-$37.27        & 11.46    & 0.12               \\
  SDSSJ0959$+$0410 & 0.126 & 10.91 & 3.41    & 196.93 & $-$36.51        & 10.93    & 0.18               \\
  SDSSJ1016$+$3859 & 0.168 & 11.23 & 4.73    & 245.83 & $-$36.77        & 11.27    & 0.20               \\
  SDSSJ1020$+$1122 & 0.282 & 11.54 & 6.46    & 281.81 & $-$37.13        & 11.52    & 0.11               \\
  SDSSJ1023$+$4230 & 0.191 & 11.33 & 6.57    & 238.13 & $-$37.03        & 11.38    & 0.23               \\
  SDSSJ1029$+$0420 & 0.104 & 11.04 & 4.11    & 206.25 & $-$36.68        & 11.05    & 0.16               \\
  SDSSJ1032$+$5322 & 0.133 & 10.90 & 2.81    & 299.45 & $-$35.93        & 11.21    & 0.59               \\
  SDSSJ1100$+$5329 & 0.317 & 11.59 & 13.64   & 180.75 & $-$38.01        & 11.46    & $-$0.11              \\
  SDSSJ1103$+$5322 & 0.158 & 11.29 & 7.11    & 190.65 & $-$37.31        & 11.22    & 0.02               \\
  SDSSJ1106$+$5228 & 0.095 & 11.13 & 4.48    & 255.19 & $-$36.55        & 11.27    & 0.37               \\
  SDSSJ1112$+$0826 & 0.273 & 11.48 & 7.60    & 316.83 & $-$36.92        & 11.69    & 0.46               \\
  SDSSJ1134$+$6027 & 0.153 & 11.26 & 5.71    & 234.74 & $-$36.91        & 11.31    & 0.22               \\
  SDSSJ1142$+$1001 & 0.222 & 11.30 & 7.52    & 217.26 & $-$37.13        & 11.36    & 0.24               \\
  SDSSJ1143$-$0144 & 0.106 & 11.36 & 10.51   & 252.29 & $-$37.12        & 11.63    & 0.54               \\
  SDSSJ1153$+$4612 & 0.180 & 11.08 & 4.33    & 226.54 & $-$36.62        & 11.16    & 0.27               \\
  SDSSJ1204$+$0358 & 0.164 & 11.20 & 4.63    & 265.73 & $-$36.61        & 11.33    & 0.34               \\
  SDSSJ1205$+$4910 & 0.215 & 11.48 & 9.01    & 273.43 & $-$37.14        & 11.64    & 0.39               \\
  SDSSJ1213$+$6708 & 0.123 & 11.24 & 7.42    & 280.46 & $-$36.74        & 11.58    & 0.62               \\
  SDSSJ1218$+$0830 & 0.135 & 11.35 & 9.01    & 209.14 & $-$37.30        & 11.41    & 0.24               \\
  SDSSJ1250$+$0523 & 0.232 & 11.53 & 7.04    & 248.97 & $-$37.32        & 11.45    & 0.00               \\
  SDSSJ1251$-$0208 & 0.224 & 11.26 & 19.16   & 218.67 & $-$37.36        & 11.77    & 0.83               \\
  SDSSJ1306$+$0600 & 0.173 & 11.19 & 6.86    & 231.79 & $-$36.87        & 11.38    & 0.43               \\
  SDSSJ1313$+$4615 & 0.185 & 11.33 & 6.65    & 258.31 & $-$36.92        & 11.46    & 0.34               \\
  SDSSJ1318$-$0313 & 0.240 & 11.43 & 15.80   & 202.36 & $-$37.59        & 11.62    & 0.43               \\
  SDSSJ1330$-$0148 & 0.081 & 10.43 & 2.08    & 185.91 & $-$35.76        & 10.67    & 0.49               \\
  SDSSJ1402$+$6321 & 0.205 & 11.55 & 8.78    & 259.66 & $-$37.31        & 11.58    & 0.20               \\
  SDSSJ1403$+$0006 & 0.189 & 11.20 & 5.88    & 210.66 & $-$36.97        & 11.23    & 0.19               \\
  SDSSJ1416$+$5136 & 0.299 & 11.40 & 6.11    & 241.00 & $-$37.12        & 11.36    & 0.07               \\
  SDSSJ1420$+$6019 & 0.063 & 10.93 & 2.58    & 201.48 & $-$36.49        & 10.83    & $-$0.05              \\
  SDSSJ1430$+$4105 & 0.285 & 11.68 & 11.60   & 312.62 & $-$37.32        & 11.87    & 0.43               \\
  SDSSJ1432$+$6317 & 0.123 & 11.46 & 13.72   & 185.35 & $-$37.72        & 11.48    & 0.18               \\
  SDSSJ1436$-$0000 & 0.285 & 11.45 & 12.58   & 216.59 & $-$37.47        & 11.58    & 0.35               \\
  SDSSJ1443$+$0304 & 0.134 & 10.87 & 3.28    & 209.87 & $-$36.35        & 10.97    & 0.30               \\
  SDSSJ1451$-$0239 & 0.125 & 11.17 & 5.91    & 216.79 & $-$36.89        & 11.25    & 0.28               \\
  SDSSJ1525$+$3327 & 0.358 & 11.78 & 16.99   & 253.42 & $-$37.83        & 11.85    & 0.26               \\
  SDSSJ1531$-$0105 & 0.160 & 11.43 & 9.14    & 268.13 & $-$37.10        & 11.63    & 0.44               \\
  SDSSJ1538$+$5817 & 0.143 & 11.03 & 3.99    & 188.45 & $-$36.80        & 10.96    & 0.02               \\
  SDSSJ1614$+$4522 & 0.178 & 11.21 & 8.79    & 176.01 & $-$37.31        & 11.25    & 0.20               \\
  SDSSJ1621$+$3931 & 0.245 & 11.45 & 10.63   & 228.87 & $-$37.36        & 11.56    & 0.31               \\
  SDSSJ1627$-$0053 & 0.208 & 11.45 & 6.95    & 285.51 & $-$36.98        & 11.56    & 0.32               \\
  SDSSJ1630$+$4520 & 0.248 & 11.61 & 7.94    & 271.72 & $-$37.33        & 11.58    & 0.09               \\
  SDSSJ1636$+$4707 & 0.228 & 11.38 & 6.69    & 228.65 & $-$37.17        & 11.35    & 0.10               \\
  SDSSJ1644$+$2625 & 0.137 & 11.18 & 5.62    & 224.06 & $-$36.85        & 11.26    & 0.27               \\
  SDSSJ1719$+$2939 & 0.181 & 11.22 & 5.56    & 283.17 & $-$36.61        & 11.46    & 0.50               \\
  SDSSJ2238$-$0754 & 0.137 & 11.20 & 5.83    & 193.36 & $-$37.10        & 11.15    & 0.05               \\
  SDSSJ2300$+$0022 & 0.228 & 11.40 & 7.02    & 275.51 & $-$36.96        & 11.54    & 0.36               \\
  SDSSJ2303$+$1422 & 0.155 & 11.47 & 9.50    & 244.28 & $-$37.28        & 11.56    & 0.29               \\
  SDSSJ2321$-$0939 & 0.082 & 11.35 & 7.42    & 234.96 & $-$37.10        & 11.42    & 0.26               \\
  SDSSJ2341$+$0000 & 0.186 & 11.48 & 14.04   & 195.89 & $-$37.67        & 11.54    & 0.25               \\
  SDSSJ2347$-$0005 & 0.417 & 11.58 & 9.90    & 400.33 & $-$36.87        & 12.01    & 0.74               \\
  average          & 0.211 & 11.40 & 8.50    & 243.24 & $-$36.85        & 11.55    & 0.36               \\
\end{longtable}
\endgroup

\normalem
\begin{acknowledgements}
We thank the anonymous referee for their valuable suggestions, which helped improve the manuscript. This work is supported by the NSFC grant (No 11988101) and the K.C.Wong Education Foundation. 
\end{acknowledgements}

\bibliographystyle{raa}

\end{document}